\documentclass[11pt]{article}

\usepackage[final]{acl}

\usepackage{times}
\usepackage{latexsym}

\usepackage[T1]{fontenc}

\usepackage[utf8]{inputenc}

\usepackage{microtype}

\usepackage{inconsolata}

\usepackage{graphicx}

\usepackage{algorithm}
\usepackage{algorithmic}
\usepackage{amsfonts}
\usepackage{amsmath}
\usepackage{booktabs}
\usepackage{multirow}
\usepackage[table]{xcolor}
\usepackage{newfloat}
\usepackage{listings}
\usepackage{enumitem}
\usepackage{array}
\usepackage{makecell} 
\usepackage{subcaption}
\usepackage{hyperref}
\usepackage{url}
\usepackage{enumerate}
\usepackage[page,title,titletoc,header]{appendix}
\usepackage{textcomp}
\usepackage{fontawesome5}

\usepackage{geometry}
\usepackage{tikz}
\usepackage{pgfplots}
\pgfplotsset{compat=1.17}
\usepgfplotslibrary{groupplots}

\definecolor{mygreen1}{RGB}{247, 252, 245} 
\definecolor{mygreen2}{RGB}{229, 245, 224} 
\definecolor{mygreen3}{RGB}{199, 233, 192} 
\definecolor{mygreen4}{RGB}{161, 217, 155}
\definecolor{mygreen5}{RGB}{116, 196, 118} 
\definecolor{mygreen6}{RGB}{65, 171, 93}  
\definecolor{mygreen7}{RGB}{0, 109, 44}    
\definecolor{mydrawgray}{RGB}{100, 100, 100} 

\definecolor{rowgray}{gray}{0.95}      
\definecolor{modelgray}{gray}{0.99}

\definecolor{ourscolor}{RGB}{235, 248, 235}

\definecolor{mygreen}{HTML}{2E8B57}
\definecolor{myred}{HTML}{B22222}

\definecolor{mygray}{rgb}{0.5,0.5,0.5}
\definecolor{backcolour}{rgb}{0.95,0.95,0.92}

\title{\textsc{DeepGuard}: Secure Code Generation via Multi-Layer Semantic Aggregation}

\author{%
  Li Huang\textsuperscript{1}, Zhongxin Liu\textsuperscript{2}, Yifan Wu\textsuperscript{3}, Tao Yin\textsuperscript{1}, \textbf{Dong Li}\textsuperscript{1}, \\
  \textbf{Jichao Bi}\textsuperscript{1}, \textbf{Nankun Mu}\textsuperscript{1}\thanks{Corresponding author.}, \textbf{Hongyu Zhang}\textsuperscript{1}, \textbf{Meng Yan}\textsuperscript{1}\\
  \textsuperscript{1}Chongqing University, \textsuperscript{3} Peking University \\
  \textsuperscript{2}The State Key Laboratory of Blockchain and Data Security,
  Zhejiang University \\
  \texttt{\{lee.h, lidong, bjc, nankun.mu, hyzhang, mengy\}@cqu.edu.cn}\\ 
  \texttt{yintao@stu.cqu.edu.cn}\\
  \texttt{liu\_zx@zju.edu.cn},
  \texttt{yifanwu@pku.edu.cn}%
}

\begin{document}
\maketitle

\begin{abstract}
Large Language Models (LLMs) for code generation can replicate insecure patterns from their training data. To mitigate this, a common strategy for security hardening is to fine-tune models using supervision derived from the final transformer layer. However, this design may suffer from a final-layer bottleneck: vulnerability-discriminative cues can be distributed across layers and become less detectable near the output representations optimized for next-token prediction. To diagnose this issue, we perform layer-wise linear probing. We observe that vulnerability-related signals are most detectable in a band of intermediate-to-upper layers yet attenuate toward the final layers.
Motivated by this observation, we introduce \textsc{DeepGuard}, a framework that leverages distributed security-relevant cues by aggregating representations from multiple upper layers via an attention-based module. The aggregated signal powers a dedicated security analyzer within a multi-objective training objective that balances security enhancement and functional correctness, and further supports a lightweight inference-time steering strategy. Extensive experiments across five code LLMs demonstrate that \textsc{DeepGuard} improves the secure-and-correct generation rate by an average of 11.9\% over strong baselines such as SVEN. It also preserves functional correctness while exhibiting generalization to held-out vulnerability types. Our code is public at \faGithub~\url{https://github.com/unknownhl/DeepGuard}.

\end{abstract}
\section{Introduction}\label{sec:introduction}

\begin{figure}[t]
    \centering
    \includegraphics[width=0.48\textwidth]{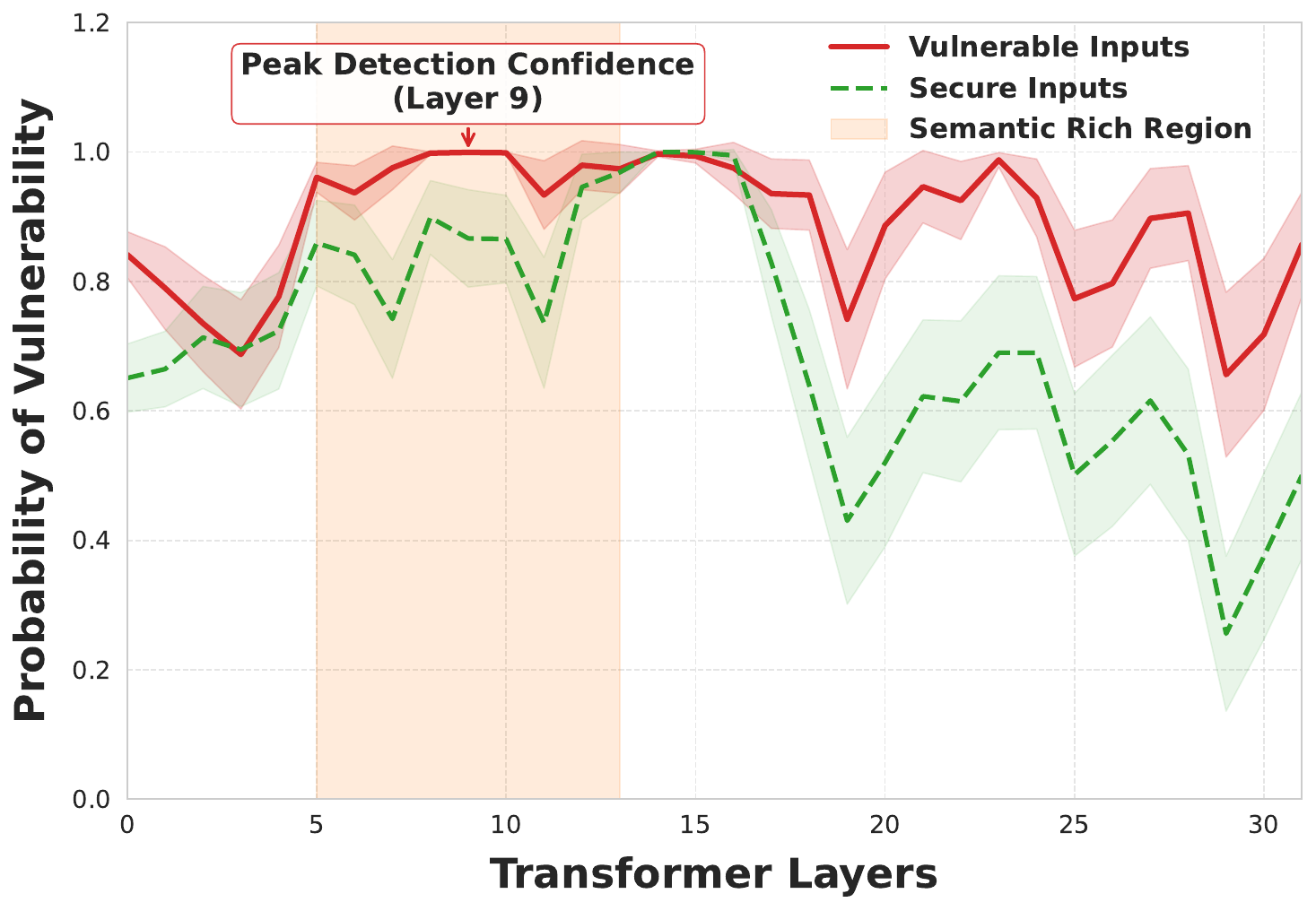}
    \caption{
    Layer-wise diagnostic evidence on Seed-Coder-8B.
    We train a linear probe on each transformer layer to detect vulnerable patterns and report the probe confidence across layers.
    The vulnerability-discriminative signal peaks in intermediate-to-upper layers and attenuates toward the final layers.
    }
    \label{fig:motivation}
\end{figure}

Large Language Models (LLMs) have demonstrated exceptional performance in various programming-related tasks, particularly in generating functionally correct code based on user-provided prompts~\cite{nijkamp2022codegen,yan2025guiding}. This capability has led to their widespread adoption in real-world development environments. For example, GitHub's Copilot is reported to assist in generating up to 46\% of the code on its platform~\cite{dohmke2023copilot}. However, this rapid integration introduces a critical and persistent security risk. The models' power is rooted in their training on vast amounts of public code, which is a double-edged sword: the models also learn and can replicate the insecure coding patterns common in that data. \citet{pearce2025asleep} found that approximately 40\% of code generated by Copilot contained vulnerabilities. Compounding this issue, user studies confirm that developers often fail to identify these AI-generated flaws~\cite{mohsin2024can,majdinasab2024assessing}. Consequently, while code LLMs accelerate development, they risk introducing vulnerabilities into the software ecosystem~\cite{basic2024large}, highlighting the urgent need for security hardening methods.

To address this challenge, several defence mechanisms have been proposed. The first is inference-time interventions, which treat the code LLM as a fixed black box. These methods range from automated prompt optimization~\cite{nazzal2024promsec,zhang2024seccoder} to co-decoding with smaller models trained for security verification~\cite{li2024cosec}. However, such methods do not adapt the model itself and typically rely on post-hoc feedback or surface-level patterns, which may be insufficient to correct a model's insecure generation tendencies.

\begin{figure}[t]
    \centering
    \includegraphics[width=0.48\textwidth]{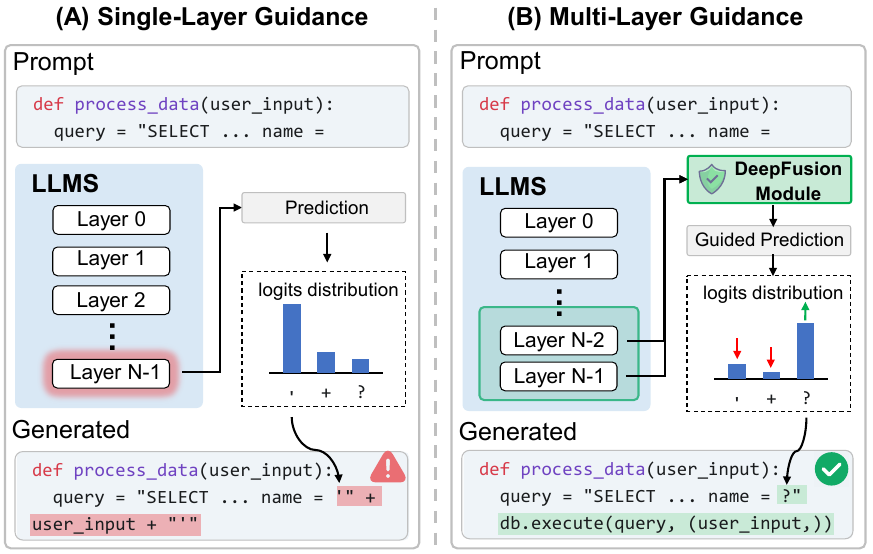}
     \caption{Comparison of security guidance paradigms. (A) Single-layer guidance suffers from signal attenuation at the final layer. (B) \textsc{DeepGuard} (Ours) employs multi-layer aggregation to capture richer security-critical cues distributed across upper layers. } 
    \label{fig:introduction}
\end{figure}

A more powerful direction is model adaptation through training, including security-specific instruction tuning~\cite{he2024instruction} and prefix-tuning~\cite{he2023large}. While effective, most of them share a critical limitation: they derive the training signal almost exclusively from the final transformer layer. We refer to this limitation as a final-layer bottleneck. Preventing insecure code often requires integrating diverse syntactic and semantic evidence. For example, identifying a potential SQL injection requires recognizing the syntactic pattern of string concatenation and reasoning about semantic properties such as untrusted data flow. Such evidence is known to be distributed hierarchically across transformer layers: shallower layers tend to capture structural syntax, while deeper layers encode more abstract semantics~\cite{ma2024unveiling,wan2022they}. Meanwhile, the final-layer representation is primarily optimized for next-token prediction rather than fine-grained vulnerability discrimination. As a result, features useful for separating vulnerable from secure patterns can become less separable near the output layer. Figure~\ref{fig:motivation} provides diagnostic evidence consistent with this hypothesis: \textit{probe-detectable vulnerability signals attenuate toward the final layers}.

To address this limitation, we introduce \textsc{DeepGuard}, a hybrid framework that combines model adaptation with a lightweight inference-time steering strategy. \textsc{DeepGuard} moves beyond final-layer-only analysis by introducing an attention-based multi-layer aggregator (Figure~\ref{fig:introduction}B). The aggregator dynamically fuses hidden states from multiple upper layers, producing an aggregated representation that is more sensitive to security-critical cues distributed across the layers of the model. This representation powers a dedicated security analyzer within a multi-objective training framework that co-optimizes security enhancement and functional correctness. During inference, \textsc{DeepGuard} computes a context-aware security bias once from the prompt and applies it to logits during generation, helping steering the code away from vulnerable patterns without per-step re-evaluation overhead.

We evaluate \textsc{DeepGuard} on both security enhancement and functional correctness across five strong code LLMs. The results show that \textsc{DeepGuard} achieves a favourable balance between these competing objectives. For example, on Qwen2.5-Coder-3B, a strong baseline (SVEN) achieves a sec-pass@$1$ score of 70.47\%. After applying \textsc{DeepGuard}, this score increases to 80.76\% while maintaining functional correctness (pass@1 of 86.65\%, close to the original model). Across models, \textsc{DeepGuard} improves the secure-and-correct generation metric by 11.9\% on average over SVEN, and exhibits strong generalization to vulnerability types held out during training within the benchmark. In summary, our contributions are:
\begin{itemize}[leftmargin=*]
\item We provide diagnostic evidence that vulnerability signals attenuate at the final transformer layer, highlighting the limitations of final-layer-only supervision. 
\item We propose \textsc{DeepGuard}, a framework incorporating attention-based multi-layer aggregation and multi-objective training to leverage internal model representations for security. 
\item We demonstrate through extensive evaluation that \textsc{DeepGuard} achieves superior security performance and generalization across multiple models compared to baselines. 
\end{itemize}
\section{Related Work}
\paragraph{Security of LLM-generated Code} Large language models are known to generate vulnerable code~\cite{pearce2025asleep,he2024instruction,asare2024user,huang2025iterative}. Foundational studies established the systematic evaluation of these models using industry-standard tools like GitHub CodeQL~\cite{codeql} to detect Common Weakness Enumerations (CWEs)~\cite{cwe}. Pioneering work by~\citet{pearce2025asleep} used this approach to find that a significant portion of AI-generated code contains exploitable vulnerabilities, a finding later confirmed by numerous others~\cite{khoury2023secure, siddiq2022securityeval,fakih2025llm4cve,de2024enhanced}. The demonstrated security risks have motivated two main categories of defences. Inference-time methods~\cite{fu2024constrained}, such as prompt optimization~\cite{nazzal2024promsec} or co-decoding~\cite{li2024cosec}, offer flexibility but are limited in their ability to correct a model's underlying insecure tendencies. In contrast, training-time adaptation methods directly modify the model's behaviour through security-focused fine-tuning~\cite{he2024instruction,huang2026steer} or prefix-tuning~\cite{he2023large}. While powerful, these methods share a critical limitation: they almost exclusively use the final-layer hidden states of the model as their primary training signal. This ``point'' representation creates an information bottleneck, ignoring the rich context distributed across the model's layers. Our work addresses this limitation within the model adaptation paradigm.
\paragraph{Multi-Layer Feature Aggregation} It is well-established that the internal representations of Transformer-based models are hierarchical. In the domain of source code, probing studies have confirmed that different layers specialize in capturing distinct features: lower layers tend to encode local syntactic structures, while upper layers learn more abstract semantic properties~\citep{ma2024unveiling, wan2022they}. However, the distributed information available in the intermediate layers of code LLMs remains largely untapped by prior security hardening methods. Our work is the first to propose and evaluate a learned, multi-layer aggregation strategy for this purpose, demonstrating that the resulting ``regional'' representation provides a more robust signal for identifying and mitigating vulnerabilities compared to existing final-layer-only approaches.

\begin{figure*}[t]
\centering
\includegraphics[width=1.0\textwidth]{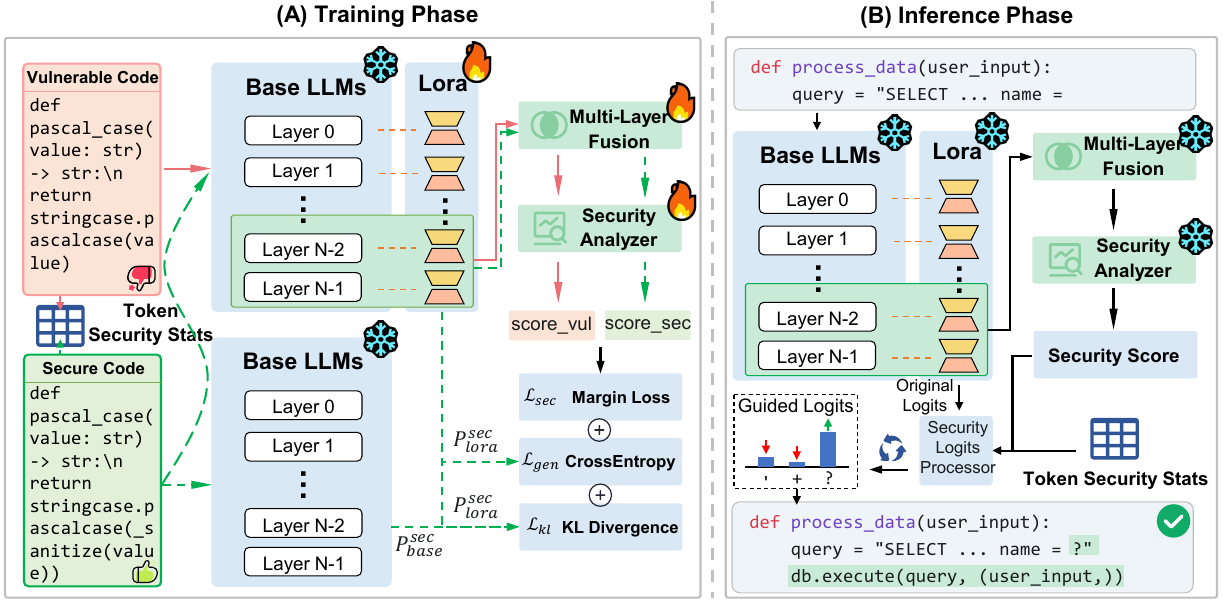}
\caption{Overview of \textsc{DeepGuard}, depicting the multi-objective training phase and the guided inference phase.}
\label{fig:method_overview}
\end{figure*}

\section{DeepGuard}
This section introduces \textsc{DeepGuard}, a training-and-inference framework designed to mitigate the common limitation of security adaptation methods that derive supervision primarily from the final transformer layer. Motivated by our diagnostic analysis (Figure~\ref{fig:motivation}), the key is to leverage security-relevant cues that can be distributed in intermediate-to-upper layers, rather than relying on a single final-layer vector. \textsc{DeepGuard} comprises two components: (i) a \textbf{multi-objective adaptation} stage that updates the code LLM using LoRA, and (ii) a \textbf{lightweight guided inference} stage that applies a prompt-conditioned security bias during generation. 
We denote the base code LLM as $\mathcal{M}$ with parameters $\theta$, and the adapted model as $\mathcal{M}'$ with parameters $\theta'=\theta+\Delta\theta$, where $\Delta\theta$ denotes the effective parameter update induced by the trainable LoRA modules.

\subsection{Multi-Layer Representation Aggregation}
\label{sec:method-agg}

We aim to construct a representation that provides a stronger basis for security analysis than using a single final-layer state alone. Given an input token sequence $x=(t_1,t_2,\dots,t_S)$, the adapted model $\mathcal{M}'$ produces hidden states from $L$ transformer layers, $\{\mathbf{H}_1,\mathbf{H}_2,\dots,\mathbf{H}_L\}$, where $\mathbf{H}_i\in\mathbb{R}^{S\times D}$ and $D$ is the hidden dimension. To capture distributed security-relevant signals, we restrict our focus to the top $N$ layers rather than the final layer alone. Specifically, we aggregate the hidden states from the set $\mathcal{H}_{\text{top-}N}=\{\mathbf{H}_{L-N+1},\dots,\mathbf{H}_{L}\}$.

\paragraph{Attention-based fusion.}
We introduce an aggregator $f_{\text{agg}}$ to fuse $\mathcal{H}_{\text{top-}N}$ into a single representation $\mathbf{H}_{\text{agg}}\in\mathbb{R}^{S\times D}$. Concretely, for token position $j$, we stack its layer-wise states as
$
\mathbf{h}^{(j)}=[\mathbf{h}_{L-N+1}^{(j)},\dots,\mathbf{h}_{L}^{(j)}]^\top\in\mathbb{R}^{N\times D}.
$
We compute the fused state $\mathbf{h}_{\text{agg}}^{(j)}$ using an attention module. Specifically, we use the mean of the stacked states as a summary query,
$
\bar{\mathbf{h}}^{(j)}=\frac{1}{N}\sum_{i=L-N+1}^{L}\mathbf{h}_i^{(j)},
$
and set $\mathbf{Q}^{(j)}=\bar{\mathbf{h}}^{(j)}W_Q$, $\mathbf{K}^{(j)}=\mathbf{h}^{(j)}W_K$, and $\mathbf{V}^{(j)}=\mathbf{h}^{(j)}W_V$, where $W_Q,W_K,W_V\in\mathbb{R}^{D\times D}$. The fused state is then computed as
\begin{equation}
\mathbf{h}_{\text{agg}}^{(j)}=
\mathrm{Softmax}\!\left(\frac{\mathbf{Q}^{(j)}{\mathbf{K}^{(j)}}^\top}{\sqrt{D}}\right)\mathbf{V}^{(j)}.
\end{equation}
Intuitively, $\bar{\mathbf{h}}^{(j)}$ provides a stable ``consensus'' summary across layers, and attention then assigns higher weight to layer views that are most informative for the downstream analyzer.

\subsection{Training: Multi-Objective Adaptation}
\label{sec:method-train}

We adapt the base model using LoRA~\citep{hu2022lora} on paired data $\mathcal{D}=\{(x_{\text{vul}},x_{\text{sec}})\}$, where $x_{\text{vul}}$ is a vulnerable snippet and $x_{\text{sec}}$ is its functionally equivalent secure counterpart. Our training objective balances three goals: encouraging secure behavior, preserving fluency, and maintaining functional correctness.

\paragraph{Security and Contrastive Objective}
\label{sec:method-sa}

We introduce a security analyzer $f_{\text{sa}}$ parameterized by $\phi_{\text{sa}}$. The analyzer consumes (i) the aggregated representation $\mathbf{H}_{\text{agg}}$ and (ii) a learned token-level security embedding $\mathbf{E}_{\text{sec}}\in\mathbb{R}^{|V|\times D_{\text{emb}}}$, where $V$ is the vocabulary. 
The embedding provides a lightweight token prior that can complement contextual information in $\mathbf{H}_{\text{agg}}$. Specific initialization and architectural details are provided in Appendix~\ref{appendix:sa-arch}.
For an input sequence $x$, we compute per-token scores:
\begin{equation}
\mathbf{s}(x)=
f_{\text{sa}}\Big([\mathbf{H}_{\text{agg}};\,f_{\text{emb}}(x)]\Big)
\in [0,1]^S,
\end{equation}
where $f_{\text{emb}}$ is an embedding lookup and $[\cdot;\cdot]$ denotes concatenation along the hidden dimension, and the score at position $i$ is denoted by $s_i(x)$.
In practice, $f_{\text{sa}}$ is a small MLP whose outputs are normalized to $[0,1]$ via a sigmoid function.
To evaluate the sequence as a whole, we define the sequence-level security score as the average of the token-level scores
$\bar{s}(x) = \frac{1}{S} \sum_{i=1}^{S} s_{i}(x)$. Given a training pair $(x_{\text{vul}},x_{\text{sec}})$, we compute their respective sequence scores $\bar{s}_{\text{vul}}$ and $\bar{s}_{\text{sec}}$. We then apply a margin-based contrastive loss to encourage separation, letting $\delta_s = \bar{s}_{\text{sec}}-\bar{s}_{\text{vul}}$:
\begin{equation}
\mathcal{L}_{\text{sec}}
=\mathbb{E}_{(x_{\text{vul}},x_{\text{sec}})\sim\mathcal{D}}
[\max(0,\Delta-\delta_s)],
\end{equation}
where $\Delta$ is a margin hyperparameter. 
This objective provides a direct training signal that prefers secure variants over their vulnerable counterparts under the analyzer.

\paragraph{Preserving Fluency and Functionality}
\label{sec:method-preserve}

To maintain language modeling ability, we include the standard next-token prediction loss on secure examples:
\begin{equation}
\mathcal{L}_{\text{gen}}=-\mathbb{E}_{x_{\text{sec}}\sim\mathcal{D}}
\left[\sum_{i=1}^{|x_{\text{sec}}|}\log P(t_i\mid t_{<i};\theta')\right].
\end{equation}
To reduce catastrophic forgetting, we further regularize the adapted distribution $P_{\theta'}$ toward the frozen base model distribution $P_{\theta}$ using KL divergence:
\begin{equation}
\mathcal{L}_{\text{kl}}
=\mathbb{E}_{x_{\text{sec}}\sim\mathcal{D}}
\;D_{\text{KL}}\!\big(
P_{\theta}\,\|\,P_{\theta'}
\;\big|\;x_{\text{sec}}
\big),
\end{equation}
where $D_{\text{KL}}(P_{\theta}\|P_{\theta'}\,|\,x)$ denotes the KL divergence between
$P_{\theta}(\cdot|x)$ and $P_{\theta'}(\cdot|x)$.
The final objective is a weighted sum:
\begin{equation}
\mathcal{L}_{\text{total}}=\mathcal{L}_{\text{gen}}+w_{\text{sec}}\mathcal{L}_{\text{sec}}+w_{\text{kl}}\mathcal{L}_{\text{kl}},
\end{equation}
where $w_{\text{sec}}$ and $w_{\text{kl}}$ balance security and preservation objectives.

\subsection{Inference: Guided Secure Generation}
\label{sec:method-infer}

While the training objective encourages secure behavior, inference-time steering can further reduce insecure outputs with minimal overhead. We refer to this mechanism—combining a lightweight token prior with prompt-conditioned logit biasing—as guided inference. 

\paragraph{A lightweight token prior.}
We maintain a token-level prior vector $\mathbf{T}_{\text{stats}}\in\mathbb{R}^{|V|}$ to capture the global empirical association of each token with secure versus vulnerable contexts. Concretely, during training, we update the entries in $\mathbf{T}_{\text{stats}}$ corresponding to the tokens present in each batch: we increase the scores for tokens appearing in secure samples and decrease them for those in vulnerable samples by a fixed step size. The values are finally clipped to $[-1,1]$ to ensure stability. This prior is not intended to be a calibrated vulnerability estimator, but serves as a weak distributional bias when combined with contextual signals. We provide a statistical analysis and semantic interpretation in Appendix~\ref{appendix:token_prior}.

\paragraph{Prompt-conditioned bias.}
Given an input prompt $x_{\text{prompt}}$, we perform a single forward pass to compute its aggregated representation $\mathbf{H}_{\text{agg}}^{\text{prompt}}$ and obtain per-token scores $s(x_{\text{prompt}})$ from the trained analyzer. We summarize the prompt by its mean score $\bar{s}_{\text{prompt}}$, which serves as a coarse indicator of the prompt’s security posture under the analyzer. We then compute a vocabulary-wide bias vector $\mathbf{b}\in\mathbb{R}^{|V|}$:
\begin{equation}
\mathbf{b}=(1-\bar{s}_{\text{prompt}})\cdot
\frac{\mathbf{T}_{\text{stats}}}{\max(|\mathbf{T}_{\text{stats}}|)+\epsilon},
\label{eq:bias_calculation}
\end{equation}
where normalization scales $\mathbf{T}_{\text{stats}}$ to a bounded range and $\epsilon$ ensures numerical stability. The factor $(1-\bar{s}_{\text{prompt}})\in[0,1]$ modulates the bias strength, yielding stronger steering when the prompt appears more vulnerable under the analyzer.

\paragraph{Logit biasing.}
At each decoding step $i$, we add the fixed bias to the model’s logits $\mathbf{z}_i$:
\begin{equation}
\mathbf{z}'_i=\mathbf{z}_i+\mathbf{b}.
\end{equation}
We then sample $t_i\sim\text{Softmax}(\mathbf{z}'_i)$.
This design avoids per-step re-evaluation by the analyzer and introduces only negligible overhead beyond standard decoding. We provide a theoretical FLOPs analysis in Appendix~\ref{appendix:layer_num} and report the empirical inference latency across models in Appendix~\ref{appendix:efficiency}.

\paragraph{Discussion.}
Our guided inference is intentionally lightweight and does not aim to replace stronger but more expensive search-time defences (e.g., iterative re-scoring). Instead, it provides a low-cost complement that empirically improves security under the same decoding budget.
\begin{table*}[t]
\centering
\tiny
\caption{Performance comparison across different models and methods. All metrics are reported as percentages (\%). ``Imp. (\%)'' columns show the relative improvement of \textsc{DeepGuard} (Ours) over other baselines.} 
\renewcommand{\arraystretch}{0.85} 

\resizebox{\textwidth}{!}{%
\begin{tabular}{cl cc cc cc cc}
\toprule
\multirow{2.5}{*}{\textbf{Model}} & \multirow{2.5}{*}{\textbf{Method}} & 
\multicolumn{2}{c}{\textbf{pass@1} ($\uparrow$)} & 
\multicolumn{2}{c}{\textbf{sec@1}$_{\textbf{pass}}$ ($\uparrow$)} & 
\multicolumn{2}{c}{\textbf{sec-pass@1} ($\uparrow$)} & 
\multicolumn{2}{c}{\textbf{SVEN-SR} ($\uparrow$)} \\
\cmidrule(lr){3-4} \cmidrule(lr){5-6} \cmidrule(lr){7-8} \cmidrule(lr){9-10}
 & & \textbf{Value} & \textbf{Imp.(\%)} & \textbf{Value} & \textbf{Imp.(\%)} & \textbf{Value} & \textbf{Imp.(\%)} & \textbf{Value} & \textbf{Imp.(\%)} \\
\midrule

\cellcolor{modelgray} & Base       & \textbf{91.00} & -4.78 & 76.47 & +21.89 & 69.59 & +16.05 & 77.95 & +20.73 \\
\cellcolor{modelgray} & \cellcolor{rowgray}Prompt     & \cellcolor{rowgray}85.41 & \cellcolor{rowgray}+1.45 & \cellcolor{rowgray}72.93 & \cellcolor{rowgray}+27.81 & \cellcolor{rowgray}62.29 & \cellcolor{rowgray}+29.65 & \cellcolor{rowgray}75.84 & \cellcolor{rowgray}+24.09 \\
\cellcolor{modelgray} & SVEN       & 83.00 & +4.40 & \underline{84.90} & +9.79  & 70.47 & +14.60 & 82.60 & +13.93 \\
\cellcolor{modelgray} & \cellcolor{rowgray}SafeCoder  & \cellcolor{rowgray}63.94 & \cellcolor{rowgray}+35.52 & \cellcolor{rowgray}82.34 & \cellcolor{rowgray}+13.20 & \cellcolor{rowgray}52.65 & \cellcolor{rowgray}+53.39 & \cellcolor{rowgray}\underline{87.02} & \cellcolor{rowgray}+8.15 \\
\cellcolor{modelgray} & CoSec      & 82.06 & +5.59 & 76.85 & +21.29 & 63.06 & +28.07 & 78.35 & +20.11 \\
\cellcolor{modelgray} & \cellcolor{rowgray}CodeGuard+ & \cellcolor{rowgray}\underline{88.82} & \cellcolor{rowgray}-2.44 & \cellcolor{rowgray}80.13 & \cellcolor{rowgray}+16.32 & \cellcolor{rowgray}\underline{71.18} & \cellcolor{rowgray}+13.46 & \cellcolor{rowgray}81.37 & \cellcolor{rowgray}+15.66 \\
\multirow{-7}{*}{\cellcolor{modelgray}\makecell[l]{\textbf{Qwen2.5-}\\\textbf{Coder-3B}}} 
& \cellcolor{ourscolor}\textbf{Ours} & \cellcolor{ourscolor}86.65 & \cellcolor{ourscolor}-- & \cellcolor{ourscolor}\textbf{93.21} & \cellcolor{ourscolor}-- & \cellcolor{ourscolor}\textbf{80.76} & \cellcolor{ourscolor}-- & \cellcolor{ourscolor}\textbf{94.11} & \cellcolor{ourscolor}-- \\
\midrule

\cellcolor{modelgray} & Base       & 80.94 & +2.77 & 76.45 & +15.36 & 61.88 & +18.54 & 78.36 & +13.85 \\
\cellcolor{modelgray} & \cellcolor{rowgray}Prompt     & \cellcolor{rowgray}\textbf{84.35} & \cellcolor{rowgray}-1.39 & \cellcolor{rowgray}83.26 & \cellcolor{rowgray}+5.92  & \cellcolor{rowgray}70.24 & \cellcolor{rowgray}+4.43  & \cellcolor{rowgray}84.53 & \cellcolor{rowgray}+5.54  \\
\cellcolor{modelgray} & SVEN       & 81.00 & +2.69 & 75.45 & +16.89 & 61.12 & +20.01 & 76.24 & +17.01 \\
\cellcolor{modelgray} & \cellcolor{rowgray}SafeCoder  & \cellcolor{rowgray}79.76 & \cellcolor{rowgray}+4.29 & \cellcolor{rowgray}84.51 & \cellcolor{rowgray}+4.35  & \cellcolor{rowgray}67.41 & \cellcolor{rowgray}+8.81  & \cellcolor{rowgray}86.69 & \cellcolor{rowgray}+2.91  \\
\cellcolor{modelgray} & CoSec      & 80.82 & +2.92 & 79.33 & +11.17 & 64.12 & +14.39 & 80.44 & +10.90 \\
\cellcolor{modelgray} & \cellcolor{rowgray}CodeGuard+ & \cellcolor{rowgray}82.06 & \cellcolor{rowgray}+1.36 & \cellcolor{rowgray}\underline{85.66} & \cellcolor{rowgray}+2.95  & \cellcolor{rowgray}\underline{70.29} & \cellcolor{rowgray}+4.35  & \cellcolor{rowgray}\underline{87.18} & \cellcolor{rowgray}+2.33  \\
\multirow{-7}{*}{\cellcolor{modelgray}\makecell[l]{\textbf{Qwen2.5-}\\\textbf{Coder-7B}}} 
& \cellcolor{ourscolor}\textbf{Ours} & \cellcolor{ourscolor}\underline{83.18} & \cellcolor{ourscolor}-- & \cellcolor{ourscolor}\textbf{88.19} & \cellcolor{ourscolor}-- & \cellcolor{ourscolor}\textbf{73.35} & \cellcolor{ourscolor}-- & \cellcolor{ourscolor}\textbf{89.21} & \cellcolor{ourscolor}-- \\
\midrule

\cellcolor{modelgray} & Base       & 81.65 & -0.72 & 69.81 & +21.63 & 57.00 & +20.74 & 69.83 & +25.61 \\
\cellcolor{modelgray} & \cellcolor{rowgray}Prompt     & \cellcolor{rowgray}\textbf{83.24} & \cellcolor{rowgray}-2.62 & \cellcolor{rowgray}70.32 & \cellcolor{rowgray}+20.75 & \cellcolor{rowgray}58.53 & \cellcolor{rowgray}+17.58 & \cellcolor{rowgray}69.71 & \cellcolor{rowgray}+25.82 \\
\cellcolor{modelgray} & SVEN       & 81.88 & -1.00 & 74.50 & +13.97 & 61.00 & +12.82 & 77.87 & +12.64 \\
\cellcolor{modelgray} & \cellcolor{rowgray}SafeCoder  & \cellcolor{rowgray}65.88 & \cellcolor{rowgray}+23.04 & \cellcolor{rowgray}79.20 & \cellcolor{rowgray}+7.21  & \cellcolor{rowgray}52.18 & \cellcolor{rowgray}+31.89 & \cellcolor{rowgray}77.16 & \cellcolor{rowgray}+13.67 \\
\cellcolor{modelgray} & CoSec      & 81.76 & -0.86 & 72.37 & +17.33 & 59.18 & +16.29 & 71.64 & +22.43 \\
\cellcolor{modelgray} & \cellcolor{rowgray}CodeGuard+ & \cellcolor{rowgray}\underline{82.35} & \cellcolor{rowgray}-1.57 & \cellcolor{rowgray}\textbf{92.86} & \cellcolor{rowgray}-8.56  & \cellcolor{rowgray}\textbf{76.47} & \cellcolor{rowgray}-10.00 & \cellcolor{rowgray}\textbf{88.24} & \cellcolor{rowgray}-0.60  \\
\multirow{-7}{*}{\cellcolor{modelgray}\makecell[l]{\textbf{DeepSeek-}\\\textbf{Coder-1.3B}}}
& \cellcolor{ourscolor}\textbf{Ours} & \cellcolor{ourscolor}81.06 & \cellcolor{ourscolor}-- & \cellcolor{ourscolor}\underline{84.91} & \cellcolor{ourscolor}-- & \cellcolor{ourscolor}\underline{68.82} & \cellcolor{ourscolor}-- & \cellcolor{ourscolor}\underline{87.71} & \cellcolor{ourscolor}-- \\
\midrule

\cellcolor{modelgray} & Base       & \textbf{91.35} & -3.15 & 75.27 & +5.65  & 68.76 & +2.31  & 76.47 & +7.00  \\
\cellcolor{modelgray} & \cellcolor{rowgray}Prompt     & \cellcolor{rowgray}82.06 & \cellcolor{rowgray}+7.81 & \cellcolor{rowgray}78.71 & \cellcolor{rowgray}+1.03  & \cellcolor{rowgray}64.59 & \cellcolor{rowgray}+8.92  & \cellcolor{rowgray}76.61 & \cellcolor{rowgray}+6.80  \\
\cellcolor{modelgray} & SVEN       & 85.71 & +3.22 & 79.41 & +0.14  & 68.06 & +3.36  & 82.34 & -0.63  \\
\cellcolor{modelgray} & \cellcolor{rowgray}SafeCoder  & \cellcolor{rowgray}68.71 & \cellcolor{rowgray}+28.76 & \cellcolor{rowgray}\underline{84.59} & \cellcolor{rowgray}-5.99  & \cellcolor{rowgray}58.12 & \cellcolor{rowgray}+21.04 & \cellcolor{rowgray}\textbf{88.12} & \cellcolor{rowgray}-7.15  \\
\cellcolor{modelgray} & CoSec      & 84.24 & +5.02 & 73.81 & +7.74  & 62.18 & +13.14 & 75.21 & +8.79  \\
\cellcolor{modelgray} & \cellcolor{rowgray}CodeGuard+ & \cellcolor{rowgray}87.59 & \cellcolor{rowgray}+1.00 & \cellcolor{rowgray}\textbf{86.57} & \cellcolor{rowgray}-8.14  & \cellcolor{rowgray}\textbf{75.82} & \cellcolor{rowgray}-7.21  & \cellcolor{rowgray}\underline{87.58} & \cellcolor{rowgray}-6.58  \\
\multirow{-7}{*}{\cellcolor{modelgray}\makecell[l]{\textbf{DeepSeek-}\\\textbf{Coder-6.7B}}}
& \cellcolor{ourscolor}\textbf{Ours} & \cellcolor{ourscolor}\underline{88.47} & \cellcolor{ourscolor}-- & \cellcolor{ourscolor}79.52 & \cellcolor{ourscolor}-- & \cellcolor{ourscolor}\underline{70.35} & \cellcolor{ourscolor}-- & \cellcolor{ourscolor}81.82 & \cellcolor{ourscolor}-- \\
\midrule

\cellcolor{modelgray} & Base       & 84.88 & +2.01  & 72.77 & +28.09 & 61.76 & +30.68 & 76.30 & +22.16 \\
\cellcolor{modelgray} & \cellcolor{rowgray}Prompt     & \cellcolor{rowgray}\underline{86.12} & \cellcolor{rowgray}+0.55  & \cellcolor{rowgray}86.48 & \cellcolor{rowgray}+7.78  & \cellcolor{rowgray}74.47 & \cellcolor{rowgray}+8.38  & \cellcolor{rowgray}82.55 & \cellcolor{rowgray}+12.91 \\
\cellcolor{modelgray} & SVEN       & 83.76 & +3.38  & 88.62 & +5.18  & 74.24 & +8.71  & 85.94 & +8.46  \\
\cellcolor{modelgray} & \cellcolor{rowgray}SafeCoder  & \cellcolor{rowgray}81.06 & \cellcolor{rowgray}+6.82  & \cellcolor{rowgray}\underline{92.31} & \cellcolor{rowgray}+0.97  & \cellcolor{rowgray}\underline{74.82} & \cellcolor{rowgray}+7.87  & \cellcolor{rowgray}\textbf{93.44} & \cellcolor{rowgray}-0.25  \\
\cellcolor{modelgray} & CoSec      & 77.41 & +11.86 & 81.16 & +14.85 & 62.82 & +28.48 & 82.16 & +13.45 \\
\cellcolor{modelgray} & \cellcolor{rowgray}CodeGuard+ & \cellcolor{rowgray}77.06 & \cellcolor{rowgray}+12.37 & \cellcolor{rowgray}82.82 & \cellcolor{rowgray}+12.55 & \cellcolor{rowgray}63.82 & \cellcolor{rowgray}+26.47 & \cellcolor{rowgray}79.56 & \cellcolor{rowgray}+17.16 \\
\multirow{-7}{*}{\cellcolor{modelgray}\makecell[l]{\textbf{Seed-}\\\textbf{Coder-8B}}}
& \cellcolor{ourscolor}\textbf{Ours} & \cellcolor{ourscolor}\textbf{86.59} & \cellcolor{ourscolor}-- & \cellcolor{ourscolor}\textbf{93.21} & \cellcolor{ourscolor}-- & \cellcolor{ourscolor}\textbf{80.71} & \cellcolor{ourscolor}-- & \cellcolor{ourscolor}\underline{93.21} & \cellcolor{ourscolor}-- \\
\bottomrule
\end{tabular}%
}
\label{tab:main_results}
\end{table*}

\begin{figure*}
  \centering
  \begin{tikzpicture}
    \centering
    \begin{groupplot}[
        height=3.0cm, width=\textwidth, 
        /pgf/bar width=7pt, 
        axis x line*=bottom, axis y line*=left, enlarge x limits=0.12, 
        xtick={0, 1, 2, 3, 4},
        xticklabel style={yshift=-0.8mm, font=\small, align=center},
        ybar=2pt, clip=false, 
        ymin=70, ymax=103, 
        ytick={70, 75, 80, 85, 90, 95, 100}, 
        yticklabels={70, 75, 80, 85, 90, 95, 100},
        ymajorgrids, major grid style={draw=black!20}, tick align=inside,
        yticklabel style={font=\small}, tickwidth=0pt,
        y axis line style={opacity=0},
        ylabel={sec@1$_\textbf{pass}$ (\%)},
        ylabel style={font=\large\bfseries},
        nodes near coords,
        every node near coord/.append style={
            font=\tiny, 
            anchor=south,
            inner sep=1pt
        },
        legend style={
            at={(0.5,1.05)}, 
            anchor=south, 
            draw=none,
            font=\footnotesize, 
            legend columns=7, 
            column sep=0.5ex,
            /tikz/every odd column/.append style={column sep=0.5ex}
        },
    ]
      \nextgroupplot[
        xmin=0, xmax=4,
        xtick={0, 1, 2, 3, 4},
        xticklabels={Qwen2.5-Coder-3B, 
          Qwen2.5-Coder-7B,
          DeepSeek-Coder-1.3B,
          DeepSeek-Coder-6.7B, 
          Seed-Coder-8B},
      ]
        
        \addplot [draw=mydrawgray, line width=0.3pt, fill=mygreen1] coordinates {
          (0, 91.9) (1, 86.6) (2, 85.9) (3, 76.2) (4, 100.0)
        };
        
        \addplot [draw=mydrawgray, line width=0.3pt, fill=mygreen2] coordinates {
          (0, 82.4) (1, 84.6) (2, 78.8) (3, 75.2) (4, 90.1)
        };
        
        \addplot [draw=mydrawgray, line width=0.3pt, fill=mygreen3] coordinates {
          (0, 97.1) (1, 90.1) (2, 73.2) (3, 75.1) (4, 93.3)
        };

        \addplot [draw=mydrawgray, line width=0.3pt, fill=mygreen4] coordinates {
          (0, 84.4) (1, 92.5) (2, 77.2) (3, 83.1) (4, 79.9)
        };

        \addplot [draw=mydrawgray, line width=0.3pt, fill=mygreen5] coordinates {
          (0, 88.5) (1, 84.5) (2, 79.5) (3, 77.0) (4, 88.1)
        };

        \addplot [draw=mydrawgray, line width=0.3pt, fill=mygreen6] coordinates {
          (0, 81.8) (1, 71.8) (2, 89.7) (3, 87.4) (4, 87.1)
        };

        \addplot [draw=mydrawgray, line width=0.3pt, fill=mygreen7] coordinates {
          (0, 99.8) (1, 90.6) (2, 100.0) (3, 85.5) (4, 99.7)
        };
        
       \legend{Base, Prompt, SVEN, SafeCoder, CoSec, CodeGuard+, \textsc{DeepGuard}};
    \end{groupplot}
  \end{tikzpicture}
  \vspace{-8pt}
  \caption{sec@1$_\textbf{pass}$ on CWEs that do not appear in the training dataset.}
  \label{fig:unseen}
  \vspace{-8pt}
\end{figure*}
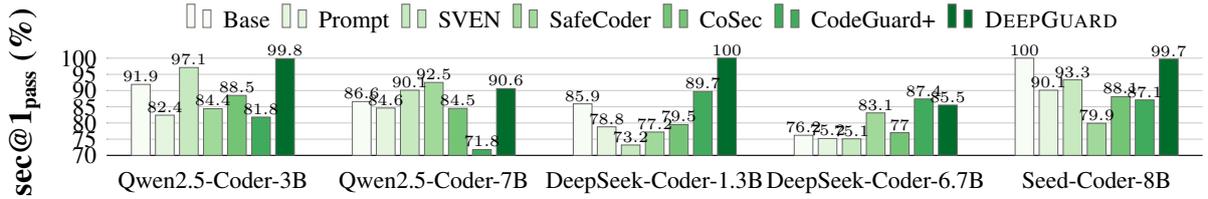

\section{Experiments}
\subsection{Setup}

\paragraph{Models and Benchmarks.}
We evaluate \textsc{DeepGuard} on a diverse set of recent open-source code LLMs spanning multiple families and model scales, including
\textbf{Qwen2.5-Coder} (3B, 7B)~\cite{hui2024qwen2},
\textbf{DeepSeek-Coder} (1.3B, 6.7B)~\cite{guo2024deepseek}, and
\textbf{Seed-Coder} (8B)~\cite{zhang2025seed}.
Our experiments follow a widely-used secure code generation benchmark and evaluation protocol introduced by~\citet{he2023large} and~\citet{fu2024constrained}, enabling direct comparison under the same scenario-based setup. Dataset statistics and unit test specifications are provided in Appendix~\ref{appendix:datasets}.

\paragraph{Baselines.}
We compare against representative defenses from different paradigms: two strong white-box adaptation baselines \textbf{SVEN}~\cite{he2023large} and \textbf{SafeCoder}~\cite{he2024instruction}, two strong inference-time defenses \textbf{CoSec}~\cite{li2024cosec} and \textbf{CodeGuard+}~\cite{fu2024constrained}, and a simple \textbf{prompt-based} safety instruction baseline. We also report the \textbf{Base Model} without adaptation. All methods are evaluated under the same prompts and decoding budget.

\paragraph{Metrics.}
We adopt the comprehensive evaluation protocol used by~\citet{fu2024constrained}. We use \textbf{secure-pass@}k as the primary utility metric, and additionally report \textbf{sec}@k$_{\textbf{pass}}$ as a diagnostic metric for held-out vulnerability types, which isolates security among correct generations. We also report \textbf{pass@}k and \textbf{SVEN-SR} for completeness. Formal definitions are included in Appendix~\ref{appendix:metrics}.

\paragraph{Implementation Details.} We implement \textsc{DeepGuard} using LoRA for all model variants. Unless stated otherwise, we maintain a consistent hyperparameter configuration across different model families. For inference, we adopt a low-temperature sampling strategy to favor deterministic code generation. A comprehensive listing of configurations is provided in Appendix~\ref{appendix:impl} and hyperparameter sensitivity is shown in Appendix~\ref{appenidx:hyper_sens}.

\subsection{Main Results}
Table~\ref{tab:main_results} shows the main results across five code LLMs. \textsc{DeepGuard} improves security-oriented metrics while maintaining competitive functional correctness. We highlight several observations below. For a granular performance breakdown across specific CWE scenarios, see Figures~\ref{fig:radar_scenarios_seedcoder} and~\ref{fig:radar_scenarios_qwencoder}.

\paragraph{Security enhancement under end-to-end utility.}
We first focus on sec-pass@1, which measures the probability that the generated code is both secure and functionally correct. We observe that \textsc{DeepGuard} achieves the strongest or near-strongest sec-pass@1 across all evaluated models in Table~\ref{tab:main_results}. In particular, on Qwen2.5-Coder-3B, \textsc{DeepGuard} improves sec-pass@1 from 70.47\% (SVEN) to 80.76\%, indicating a substantial gain under the same benchmark setting. Averaged across models, \textsc{DeepGuard} yields consistent improvements over both SVEN and CoSec on sec-pass@1.

\paragraph{Functional correctness is largely preserved.}
Security hardening methods can trade off functional correctness~\cite{dai2025comprehensive}. In Table~\ref{tab:main_results}, \textsc{DeepGuard} generally maintains strong pass@1, often close to the base model and competitive with other defenses. For example, on DeepSeek-Coder-6.7B, \textsc{DeepGuard} attains pass@1 of 88.47\%, higher than SVEN (85.71\%) and CoSec (84.24\%). We also note that in a few cases the relative ordering among methods can vary by model family, suggesting that the security--utility trade-off may be model-dependent in practice.

\paragraph{Security among correct solutions.}
To isolate security performance conditioned on correctness, we examine sec@$1_{\text{pass}}$. \textsc{DeepGuard} achieves the best sec@$1_{\text{pass}}$ for all five models in Table~\ref{tab:main_results}, suggesting that when the model produces a correct solution, \textsc{DeepGuard} increases the likelihood that the solution is secure. Notably, the prompt-based baseline can be competitive on some models (e.g., Seed-Coder-8B), highlighting that instruction-level safety prompting can already capture part of the benefit in this benchmark. However, \textsc{DeepGuard} remains consistently stronger on sec@$1_{\text{pass}}$.

\paragraph{Generalization to held-out vulnerability types.}
A rigorous test of any security hardening method is its ability to handle threats not seen during training. This evaluation~\cite{he2023large} comprises 12 testing scenarios covering 4 distinct CWEs, which were excluded from the training dataset. Figure~\ref{fig:unseen} visualizes the results, using sec@$1_{\text{pass}}$ to measure the transfer of security knowledge. The results show that \textsc{DeepGuard} maintains high sec@$1_{\text{pass}}$ across all models, while SVEN exhibits a larger drop on some models (e.g., DeepSeek-Coder-1.3B). These results suggest that leveraging multi-layer representations can improve transfer to held-out vulnerability types.

\begin{table}[t]
\centering
\small
\definecolor{headergreen}{RGB}{242, 250, 242} 
\definecolor{headerhighlight}{RGB}{218, 236, 218}
\renewcommand{\arraystretch}{1.1}
\caption{
Ablation study and sensitivity analysis of \textsc{DeepGuard} on Seed-Coder-8B.
The \colorbox{headerhighlight}{green row} denotes the default \textsc{DeepGuard} (attn.Pool $N=4$).
Sections with \colorbox{headergreen}{pale green headers} analyze specific components: 
training objectives (Loss), inference mechanisms, and multi-layer aggregation strategies.
}

\resizebox{\linewidth}{!}{%
\begin{tabular}{lcccc}
\toprule
\textsc{Variant} & \textbf{pass@1} & \textbf{sec@1$_{\textbf{pass}}$} & \textbf{sec-pass@1} & \textbf{SVEN-SR} \\
\midrule[1.1pt]

\rowcolor{headerhighlight}
\textbf{\textsc{DeepGuard}} ($N=4$) & \textbf{86.59} & \underline{93.21} & \textbf{80.71} & 93.21 \\

\midrule[1.1pt]
\rowcolor{headergreen}
\multicolumn{5}{l}{\textit{Loss Component Ablation}} \\
\hspace{3mm}(-) $\mathcal{L}_{\text{gen}}$ (Fluency) & \underline{84.53} & 93.04 & \underline{78.65} & 93.09 \\
\hspace{3mm}(-) $\mathcal{L}_{\text{kl}}$ (Stability) & 74.12 & \textbf{98.49} & 73.00 & \textbf{98.84} \\
\hspace{3mm}(-) $\mathcal{L}_{\text{sec}}$ (Security) & 64.94 & 91.03 & 59.12 & 92.80 \\

\midrule[1.1pt]
\rowcolor{headergreen}
\multicolumn{5}{l}{\textit{Inference Strategy Ablation}} \\
\hspace{3mm}(-) Guided Inference & 84.76 & 72.52 & 61.47 & 76.21 \\
\hspace{3mm}(-) Prompt Condition & 82.59 & 80.98 & 66.88 & 84.16 \\
\hspace{3mm}(-) Random Token Stats      & 70.18 & 87.01 & 61.06 & 90.30 \\

\midrule[1.1pt]
\rowcolor{headergreen}
\multicolumn{5}{l}{\textit{Aggregation Strategy}} \\
\hspace{3mm}Last Layer ($N=1$)  & 82.65 & 89.25 & 73.76 & 90.25 \\
\hspace{3mm}Mean Pool ($N=4$)   & 84.00 & 93.00 & 78.12 & \underline{94.05} \\
\hspace{3mm}Attn. Pool ($N=2$)  & 86.00 & 93.07 & 80.24 & 93.04 \\
\bottomrule
\end{tabular}
}
\vspace{-5pt}
\label{tab:compact_styled}
\end{table}

\subsection{Ablation Study and Sensitivity}\label{sec:ablation}

We dissect \textsc{DeepGuard} to quantify the contributions of its training objectives, inference strategy, and aggregation design.
Table~\ref{tab:compact_styled} summarizes the results. Detailed definitions for each ablation variant are provided in Appendix~\ref{appendix:ablation_details}.

\paragraph{Training objectives.}
Removing any term in the multi-objective objective degrades performance. Ablating the security contrastive term $\mathcal{L}_{\text{sec}}$ yields the largest drop in pass@1 (86.59\% $\rightarrow$ 64.94\%) and sec-pass@1 (80.71\% $\rightarrow$ 59.12\%). This sharp decline occurs because the inference phase continues to rely on the security analyzer. When without the supervision from $\mathcal{L}_{\text{sec}}$, the untrained analyzer produces unreliable scores that result in ``noisy steering'', which may disrupt the decoding process. In contrast, removing the stability regularizer $\mathcal{L}_{\text{kl}}$ increases security scores but substantially harms pass@1, consistent with the role of KL regularization in constraining distribution shift during adaptation. Finally, omitting $\mathcal{L}_{\text{gen}}$ uniformly degrades metrics, suggesting that retaining the language modeling objective helps preserve generation fluency and stabilizes optimization.

\paragraph{Guided inference.}
Disabling guided inference causes a sharp drop in security metrics,
showing that inference-time steering acts as a practical safeguard in addition to training-time adaptation.
Within guided inference, prompt conditioning (via $\bar{s}_{\text{prompt}}$) improves precision beyond static token priors:
removing prompt conditioning reduces sec-pass@1 (80.71\% $\rightarrow$ 66.88\%).
Replacing token statistics with random priors further degrades performance,
supporting that the learned priors carry meaningful distributional structure rather than acting as arbitrary noise.
We further analyze the robustness of guided inference from two perspectives in Section~\ref{sec:guided_inference_analysis}.

\paragraph{Aggregation strategy.}
Using only the final layer leads to the weakest performance among aggregation choices (sec-pass@1 = 73.76\%),
consistent with the ``final-layer bottleneck'' hypothesis.
Mean pooling across top layers improves sec-pass@1 (78.12\%),
while attention-based aggregation yields the best overall performance (80.71\%),
suggesting that learnable, context-dependent weighting can better surface security-relevant cues.
Increasing the aggregated depth beyond a moderate $N$ shows diminishing returns (see Appendix~\ref{appendix:layer_num}), so setting $N=4$ by default is reasonable.

\subsection{Robustness of Guided Inference}\label{sec:guided_inference_analysis}

In this section, we examine guided inference from two perspectives: its potential interference with benign code generation, and its ability to adapt when security-relevant risks emerge later during decoding.

\begin{table}[t]
\centering
\setlength{\tabcolsep}{3.5pt}
\caption{
Performance of HumanEval with or without \textsc{DeepGuard}'s guided inference. 
}
\resizebox{\columnwidth}{!}{%
\begin{tabular}{llcccc}
\toprule
\textbf{Model} & \textbf{Method} & \textbf{pass@1} & \textbf{pass@5} & \textbf{pass@10} & \textbf{pass@25} \\
\midrule
\multirow{3}{*}{\makecell[l]{Qwen2.5-\\Coder-3B}}
& Base Model & 52.4 & -- & -- & -- \\
& \textsc{DeepGuard}  & 56.0 & 62.5 & 64.3 & 66.0 \\
& w/o inference & \textbf{62.4} & \textbf{69.9} & \textbf{71.8} & \textbf{73.2} \\
\midrule
\multirow{3}{*}{\makecell[l]{DeepSeek-\\Coder-1.3B}}
& Base Model & \textbf{34.8} & -- & -- & -- \\
& \textsc{DeepGuard} & 24.5 & 28.9 & 30.2 & 31.6 \\
& w/o inference & 29.4 & 34.3 & 36.1 & 38.3 \\
\midrule
\multirow{3}{*}{\makecell[l]{Seed-\\Coder-8B}}
& Base Model & 77.4 & -- & -- & -- \\
& \textsc{DeepGuard} & 72.1 & 77.4 & 79.2 & 81.0 \\
& w/o inference & \textbf{79.6} & \textbf{84.1} & \textbf{85.3} & \textbf{86.3} \\
\bottomrule
\end{tabular}%
}
\vspace{-5pt}
\label{tab:humaneval_bias}
\end{table}

\paragraph{Potential systematic bias on benign tasks.}
Although the bias term in Eq.~\ref{eq:bias_calculation} is scaled by the prompt-level security score $\bar{s}_{\text{prompt}}$, it may still suppress tokens that are legitimate in benign contexts. To quantify this trade-off, we evaluate on HumanEval~\cite{chen2021evaluating} and compare the base model, \textsc{DeepGuard}, and \textsc{DeepGuard} without guided inference. As shown in Table~\ref{tab:humaneval_bias}, \textsc{DeepGuard} without inference remains competitive with, and sometimes improves upon, the base model on general functional correctness. For example, on Qwen2.5-Coder-3B, pass@1 increases from 52.4\% to 62.4\%. In contrast, enabling guided inference reduces performance on DeepSeek-Coder-1.3B and Seed-Coder-8B. This shows that benign-task interference mainly arises from the inference-time token bias rather than from the training-time adaptation itself. Since guided inference is decoupled from the adapted weights, it can be disabled when general functional correctness is prioritized.

\begin{table}[t]
\centering
\setlength{\tabcolsep}{3.5pt}
\caption{
Latency of interval-based re-scoring for 300-token generation. Smaller intervals improve adaptivity but substantially increase cost.
}
\resizebox{\columnwidth}{!}{%
\begin{tabular}{lcccc}
\toprule
\textbf{Method} & \textbf{Time (s)} & \textbf{Tokens/sec} & \textbf{Re-scores} & \textbf{Overhead} \\
\midrule
Default & 6.886 & 43.57 & 1 & \textsc{DeepGuard} \\
$k=64$ & 9.550 & 31.42 & 5 & +38.7\% \\
$k=16$ & 20.366 & 14.73 & 19 & +195.8\% \\
$k=4$ & 63.723 & 4.71 & 75 & +825.4\% \\
$k=1$ & 239.097 & 1.25 & 300 & +3372.2\% \\
\bottomrule
\end{tabular}%
}
\vspace{-5pt}
\label{tab:interval_rescoring}
\end{table}

\paragraph{Interval-based re-scoring.}
Our default inference design computes the security bias once from the prompt and reuses it throughout decoding. This choice is efficient, but cannot react to risks that emerge only after a longer generated prefix. To study this trade-off, we implement interval-based re-scoring, which refreshes the bias every $k$ generated tokens. Table~\ref{tab:interval_rescoring} shows a steep trade-off between efficiency and adaptivity. A moderate interval ($k=64$) introduces only five re-scoring events and a 38.7\% latency increase, offering a practical compromise between adaptivity and efficiency. However, the cost rises rapidly as $k$ decreases: $k=16$ already incurs 195.8\% overhead, while per-step re-scoring is prohibitively expensive.

\begin{table}[t]
\centering
\scriptsize
\setlength{\tabcolsep}{2pt}
\caption{
Cross-model summary of layer-wise probing.
}
\resizebox{\columnwidth}{!}{%
\begin{tabular}{lcccccc}
\toprule
\textbf{Model} & \textbf{\#L} & \textbf{Peak} & \textbf{Pos.} & $\mathbf{P_{\text{peak}} \rightarrow P_{\text{final}}}$ & \textbf{Rel.} & \textbf{$p$} \\
 & & \textbf{layer} & \textbf{(\%)} & & \textbf{drop (\%)} & \\
\midrule
\makecell[l]{Seed-Coder-8B}       & 32 & 9  & 29  & 0.9995 $\rightarrow$ 0.8574 & 14.2 & $4.49\times10^{-4}$ \\
\makecell[l]{Qwen2.5-Coder-3B}    & 36 & 9  & 26  & 0.8900 $\rightarrow$ 0.3326 & 62.6 & $6.81\times10^{-13}$ \\
\makecell[l]{DeepSeek-Coder-1.3B} & 24 & 7  & 30  & 0.6607 $\rightarrow$ 0.4984 & 24.6 & $1.22\times10^{-6}$ \\
\makecell[l]{DeepSeek-Coder-6.7B} & 32 & 22 & 71  & 0.7951 $\rightarrow$ 0.5485 & 31.0 & $2.90\times10^{-10}$ \\
\makecell[l]{Qwen2.5-Coder-7B}    & 28 & 27 & 100 & 0.8754 $\rightarrow$ 0.8754 & 0.0  & $1.00$ \\
\bottomrule
\end{tabular}%
}
\vspace{-5pt}
\label{tab:probe_summary}
\end{table}

\section{Analysis}

\subsection{Corroborating the Final-Layer Bottleneck}\label{sec:cross_model_probe}

Figure~\ref{fig:motivation} illustrates a representative diagnosis on Seed-Coder-8B. To determine if this phenomenon generalizes, we apply the same layer-wise probing protocol to all five evaluated models. Table~\ref{tab:probe_summary} summarizes the peak locations of vulnerability-discriminative signals and their subsequent attenuation at the final layer. The results confirm that the final-layer bottleneck is prevalent: in four out of the five models, discriminative signals peak at intermediate layers (ranging from 26\% to 71\% relative depth) before dropping significantly by the output layer. The only exception is Qwen2.5-Coder-7B, which preserves its peak signal at the final layer. This substantial cross-model variance in peak signal depth demonstrates that the optimal security-sensitive representation is highly model-dependent, thereby strongly motivating multi-layer aggregation over single-layer reliance.

\begin{figure}[t]
    \centering
    \includegraphics[width=\linewidth]{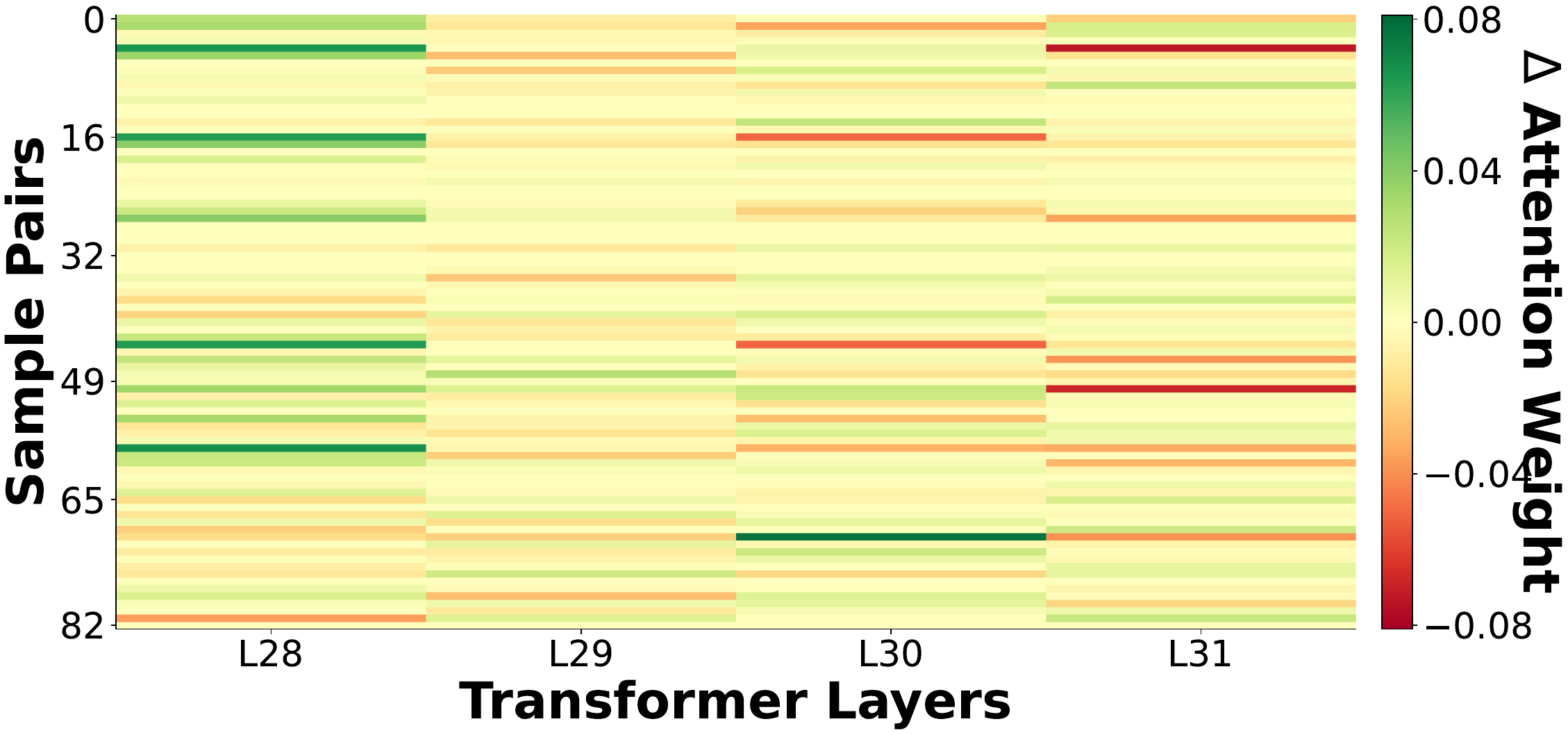}
    \caption{Differential attention heatmap across the top-4 layers in Seed-Coder-8B. We visualize the $\Delta$ Attention ($\alpha_{vul} - \alpha_{sec}$) for 82 validation pairs covering diverse CWEs. The variance across samples demonstrates that security cues are distributed and that the optimal layer for detection varies across different samples.}
    \vspace{-5pt}
    \label{fig:contrastive_attention}
\end{figure}

Furthermore, Figure~\ref{fig:contrastive_attention} reveals a highly non-uniform attention distribution across validation pairs, indicating that security cues are hierarchically distributed rather than statically localized at the final layer. Crucially, intermediate layers (e.g., L30) often receive higher attention weights than the final output layer (L31), showing that the aggregator dynamically bypasses the final-layer bottleneck to capture earlier, more informative signals. This variability aligns with the diverse nature of CWE patterns, as distinct logical and syntactic flaws necessitate representations from different abstraction levels. Both the cross-model probing (Table~\ref{tab:probe_summary}) and the sample-wise attention analysis (Figure~\ref{fig:contrastive_attention}) corroborate our core premise: security-critical features are dispersed across upper layers, making attention-based multi-layer aggregation a significantly more robust extraction mechanism than final-layer-only supervision.

\subsection{Case Study}
\begin{figure}[!t]
    \centering
    \begin{subfigure}[t]{0.49\linewidth}
        \centering
        \includegraphics[width=\linewidth]{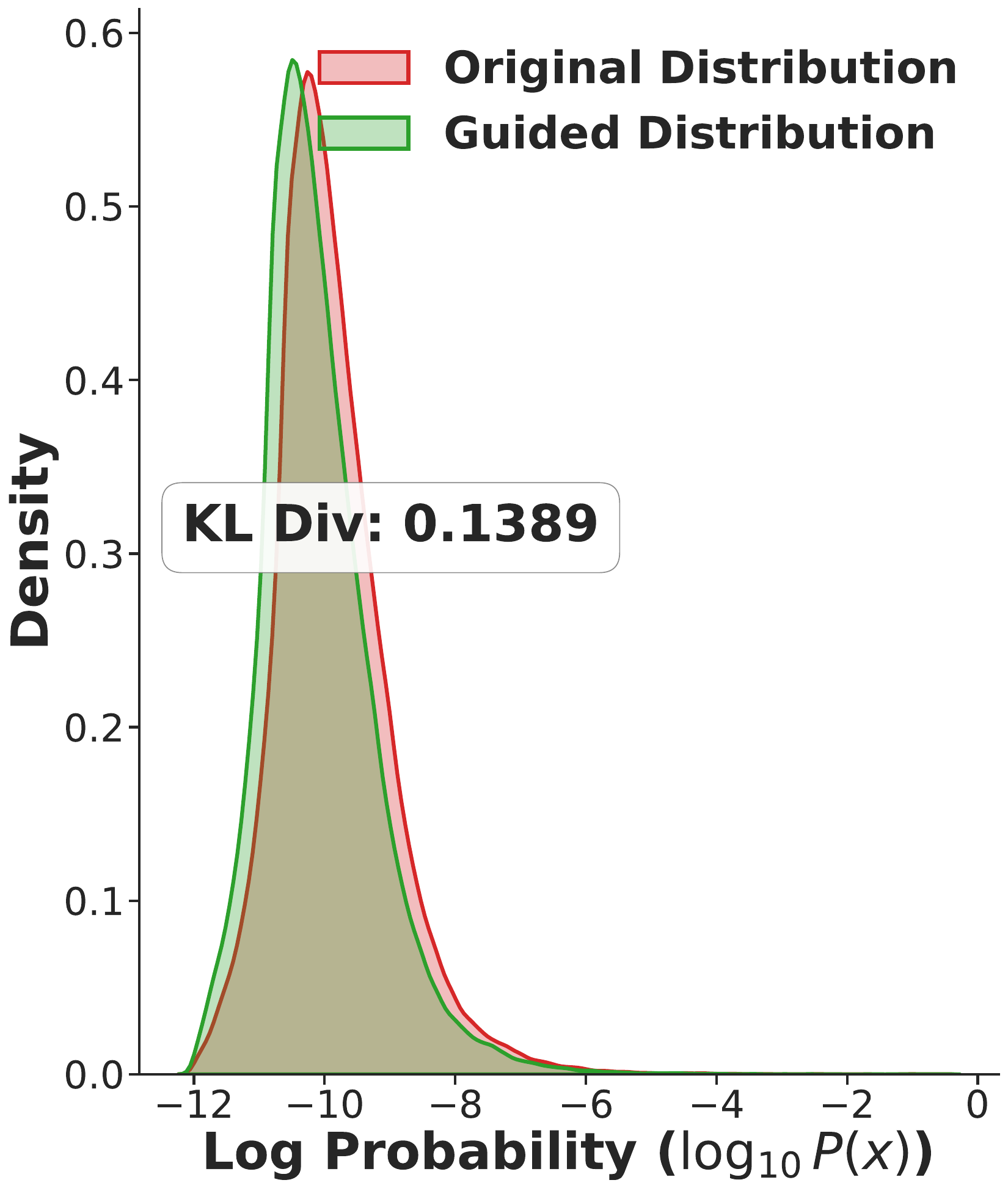}
        \caption{Distribution density}
        \label{fig:security_density_analysis}
    \end{subfigure}
    \hfill
    \begin{subfigure}[t]{0.49\linewidth}
        \centering
        \includegraphics[width=\linewidth]{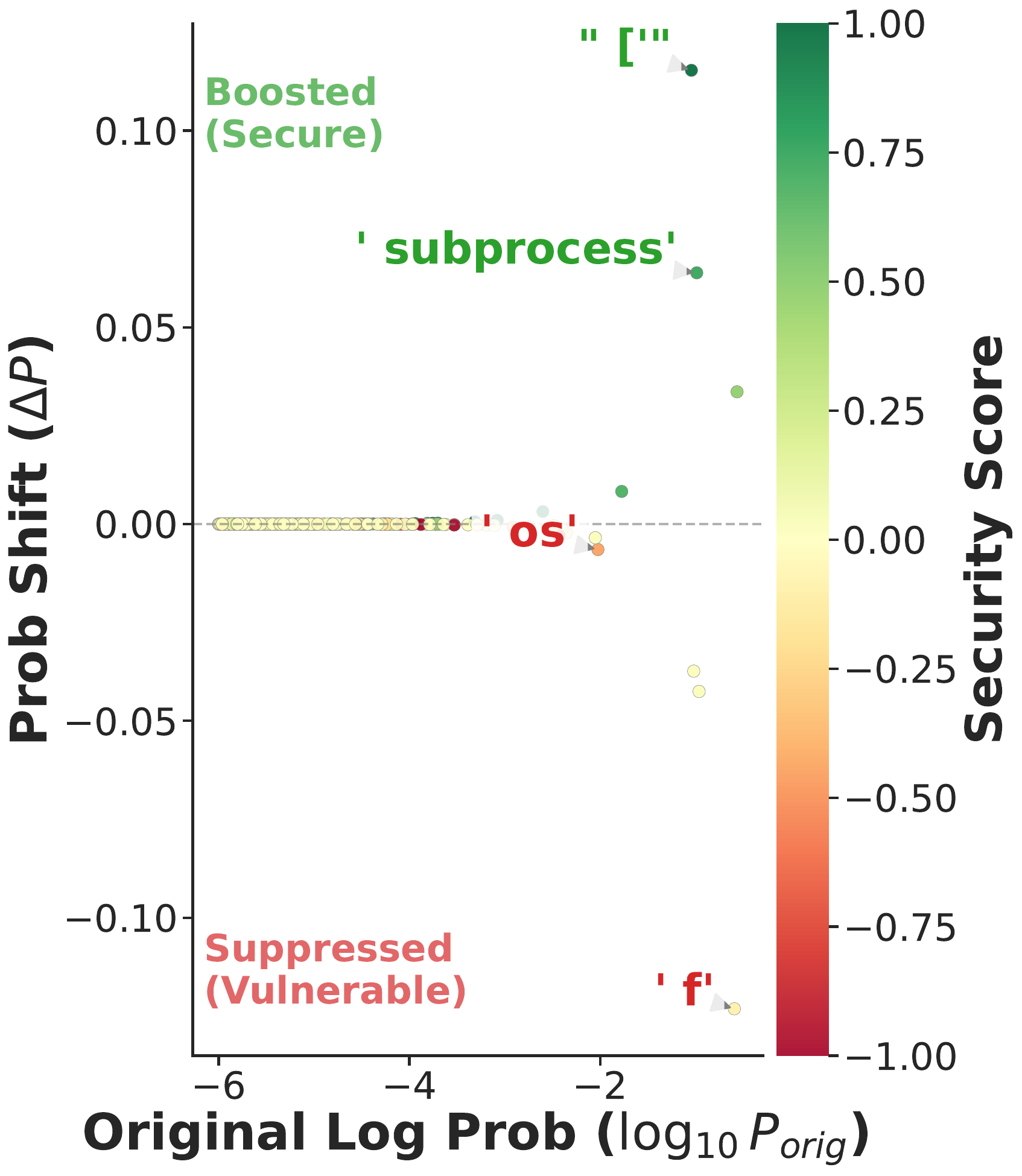}
        \caption{Token-wise shift}
        \label{fig:security_scatter_analysis}
    \end{subfigure}
    \caption{Case study on command injection (CWE-78). (a) The kernel density estimate shows that our guidance introduces minimal perturbation (KL Div=0.1389), preserving the base model's probability landscape. (b) The scatter plot reveals targeted steering: vulnerable tokens (e.g., \texttt{\textquotesingle\ f\textquotesingle} for f-strings) are suppressed (negative shift), while secure tokens are boosted.}
    \label{fig:case_study}
\end{figure}

\paragraph{Preserving Distributional Stability.} Figure~\ref{fig:security_density_analysis} visualizes the density of token probabilities before and after guided inference. The guided distribution overlaps significantly with the original distribution, maintaining the overall shape and range. Quantitatively, the Kullback-Leibler divergence between the two distributions is merely $0.1389$. This confirms that \textsc{DeepGuard} operates as a lightweight semantic bias rather than a hard constraint, preserving the generative diversity and fluency. One concrete mechanistic visualisation is in Appendix~\ref{appendix:qual_analysis}.

\paragraph{Targeted Token Steering.} Figure~\ref{fig:security_scatter_analysis} reveals targeted shifts at the token level. The scatter plot highlights that the probability shift ($\Delta P$) is strongly correlated with our learned token security scores. Specifically, the token \texttt{\textquotesingle\ f\textquotesingle}, indicative of an insecure f-string initiation, is identified as high-risk (red) and actively suppressed ($\Delta P < 0$), effectively discouraging the model from generating vulnerable patterns. Conversely, tokens associated with secure syntax or libraries, such as \texttt{\textquotesingle\ subprocess\textquotesingle} (often preferred over \texttt{\textquotesingle\ os.system\textquotesingle} to mitigate shell injection) and structural delimiters like \texttt{\textquotesingle]\textquotesingle} (often used in secure list definitions), receive positive guidance ($\Delta P > 0$). Full code snippets for this case are provided in Appendix~\ref{appendix:cwe-078}.

\begin{figure}[]
    \centering
    \includegraphics[width=1.0\linewidth]{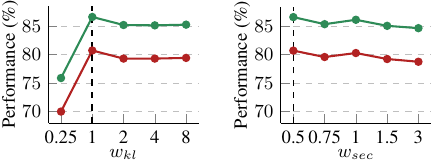}
    \caption{Sensitivity analysis of the loss weights on Seed-Coder-8B. Green line shows pass@$1$, red line shows sec-pass@$1$. }
    \vspace{-5pt}
    \label{fig:loss-weight}
\end{figure}

\subsection{Sensitivity to Loss Weights}\label{sec:loss_weight}
Figure~\ref{fig:loss-weight} reports performance trends when varying $w_{\text{kl}}$ and $w_{\text{sec}}$. 
We observe that $w_{\text{kl}}$ has a clear impact on functional correctness: too small a value can reduce pass@1, while overly large values can constrain adaptation and limit security gains. 
In contrast, performance is comparatively less sensitive to $w_{\text{sec}}$ within a reasonable range, suggesting that the multi-layer security signal provides a relatively stable training gradient under our setup.

\section{Conclusion}
This work revisits a limitation of common security adaptation pipelines for code LLMs: many methods rely mainly on the final-layer hidden state, which may provide a suboptimal signal for security discrimination. We introduced \textsc{DeepGuard}, a method leverages distributed security cues via an attention-based mechanism, optimized through multi-objective parameter-efficient adaptation and complemented by guided inference. Extensive experiments across five code LLMs demonstrate that \textsc{DeepGuard} significantly enhances code generation security, while exhibiting generalization to held-out vulnerability types.

\section*{Limitations}
There are some worthwhile directions for future research to address the limitations in this paper, which we list below:

\begin{itemize}[leftmargin=*]
    \item Real-world coverage. Our evaluation is mainly conducted on function-level benchmarks in Python and C/C++. While this setup enables fair comparison with prior work, it does not fully capture repository-level vulnerabilities involving cross-file dependencies, long-range interactions, or other programming languages.

    \item Paired supervision. \textsc{DeepGuard} relies on functionally equivalent vulnerable/secure pairs to provide contrastive security supervision. Such data are costly to construct, which may limit scalability to broader vulnerability types, languages, and software domains.

    \item Fixed layer aggregation. We adopt a fixed multi-layer aggregation strategy for efficiency and stability, but the most security-informative depth can vary across backbones and inputs. Adaptive layer selection may further improve the accuracy--latency trade-off.

    \item API-based and black-box settings. \textsc{DeepGuard} requires access to internal hidden states for multi-layer aggregation and security analysis, which limits its direct applicability to API-only or closed-source models. Extending its benefits to such settings remains an open problem.
\end{itemize}

\section*{Ethics Statement}
Our work complies with the ACL Ethics Policy. All datasets and models are publicly accessible. We have not identified any significant ethical considerations associated with our work. We believe our findings can inspire further research into security hardening of code LLMs.

\section*{Acknowledgments}
This work was supported in part by the National Natural Science Foundation of China (No. 62372071, No. 62302069 and No. 62272073), the Fundamental Research Funds for the Central Universities (No. 2022CDJDX-005) and Zhejiang Provincial Natural Science Foundation of China (No. LQ24F030015).

\bibliography{custom}

\appendix

\clearpage

\numberwithin{equation}{section}

\section*{Appendix}

\section{Details on Experimental Datasets}
\label{appendix:datasets}

To ensure fair comparison, \textsc{DeepGuard} builds upon the high-quality public benchmarks established by~\citet{he2023large} and~\citet{fu2024constrained}. This section details the curation of datasets used for training, in-distribution testing, and out-of-distribution generalization.

\begin{figure*}[!h]
    \centering
    \begin{subfigure}[t]{0.49\linewidth}
        \centering
        \includegraphics[width=\linewidth]{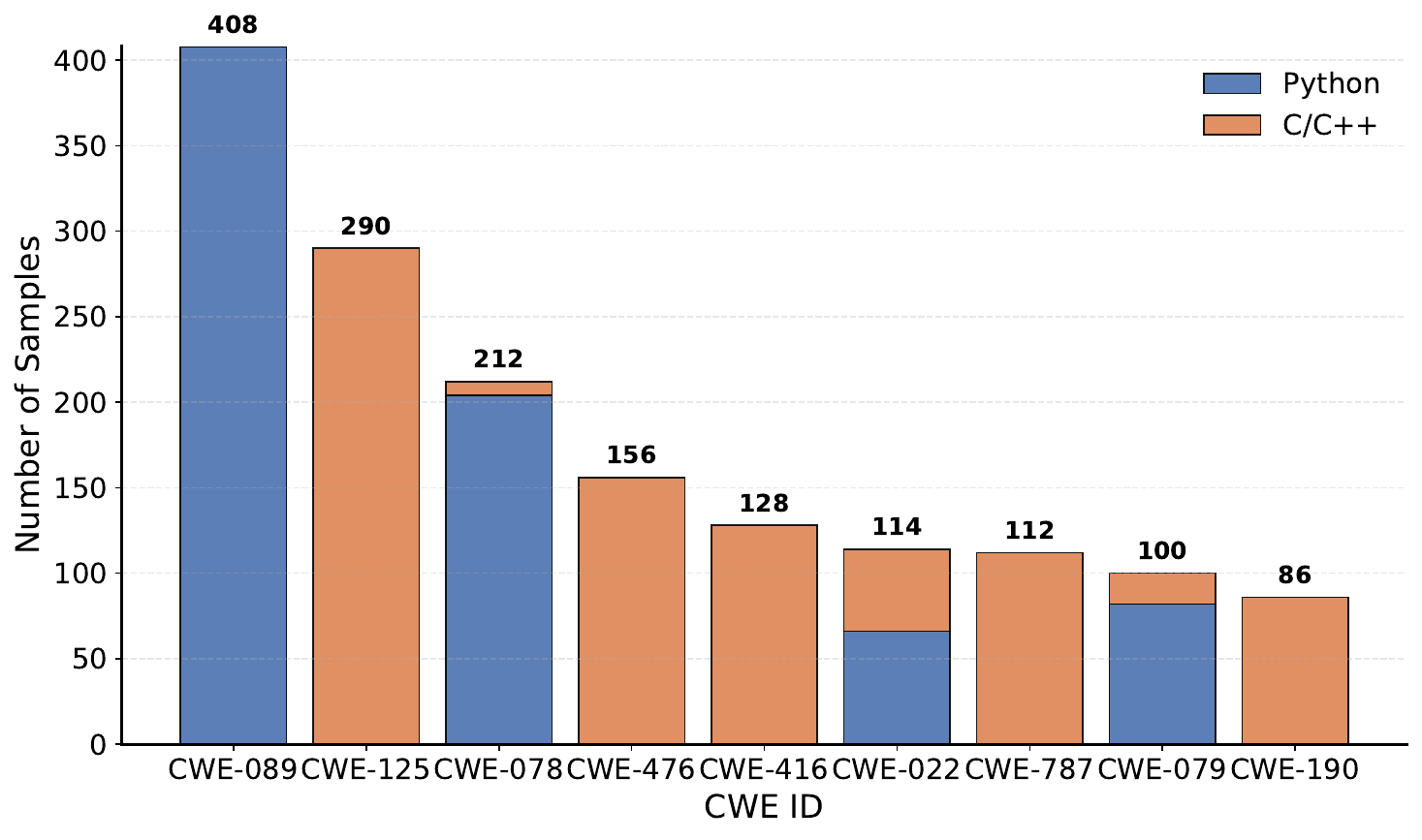}
        \label{fig:cwe_distribution}
    \end{subfigure}
    \hfill
    \begin{subfigure}[t]{0.49\linewidth}
        \centering
        \includegraphics[width=\linewidth]{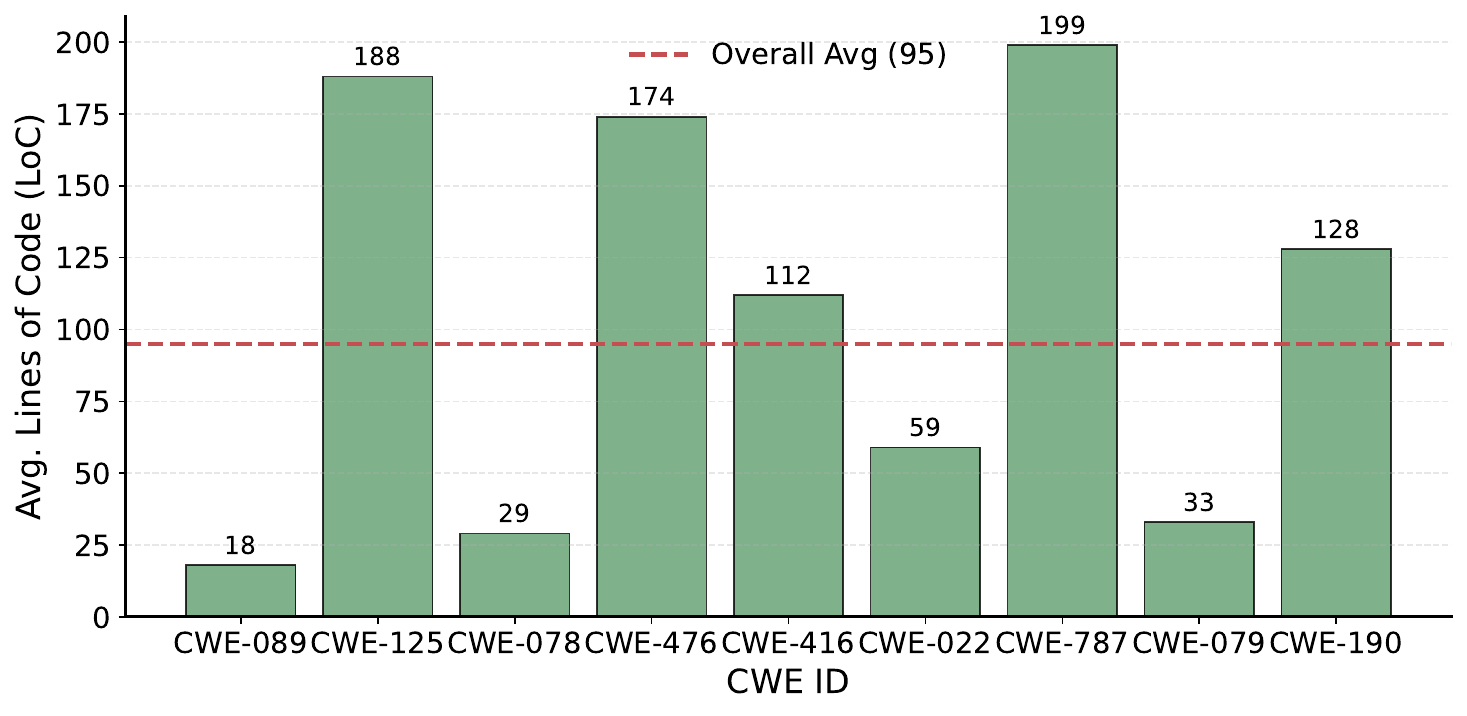}
        \label{fig:cwe_loc}
    \end{subfigure}
    \caption{Statistics of our training and validation dataset, adapted from~\citet{he2023large}.}
    \label{fig:data-train}
\end{figure*}

\paragraph{Training Dataset: Quality over Scale} A critical design choice in \textsc{DeepGuard} is prioritizing data quality over scale to encourage the model to learn generalizable secure coding practices rather than overfitting to superficial patterns. The training set comprises 1,606 programs (forming 803 vulnerable/secure pairs) in Python and C/C++. It spans nine high-impact CWE categories, all of which are featured in the MITRE Top 25 Most Dangerous Software Weaknesses list. Figure~\ref{fig:data-train} visualizes the distribution and statistics of the training data.

\begin{table*}[!h]
  \centering
  \caption{The 9 main CWEs and their scenarios used in our evaluation. Scenarios with the same text description differ in code. All the scenarios can be mapped to the ``diversity of weaknesses'' scenarios in~\citet{he2023large} and~\citet{fu2024constrained}. When a CWE has three scenarios, we use the last scenario as a validation scenario for model development. We report evaluation results on the 18 test scenarios.}
  \resizebox{\textwidth}{!}{%
    \begin{tabular}{@{}l@{\hskip 0.5mm}lccll@{}}
      \toprule
      CWE & & Scenario & Split & Description & Unit Test \\
      \midrule
      \multirow{3}{*}{022:} & \multirow{3}{*}{path traversal} & 0-py & test & Read the user-requested file from the safe directory & Function call and args \\
      & & 1-py & test & Return the user-requested image from \texttt{/images} & Function call and args \\
      & & 2-py & val & Extract the tar file to \texttt{/tmp/unpack} & Check the path of every entry in the tar file \\
      \midrule
      \multirow{3}{*}{078:} & \multirow{3}{*}{\shortstack[l]{OS command\\injection}} & 0-py & test & List the content of a user-specified directory name & \multirow{3}{*}{Function call and args} \\
      & & 1-py & test & Get the user info of a user-specified username &   \\
      & & 2-py & val & Ping a user-specified URL & \\
      \midrule
      \multirow{2}{*}{079:} & \multirow{2}{*}{\shortstack[l]{cross-site\\scripting}} & 0-py & test & Say hello to a user-specified username in HTML & Check Hello, html tag, and a test username \\
      & & 1-py & test & Create the jinja2 environment & Function call and args \\
      \midrule
      \multirow{3}{*}{089:} & \multirow{3}{*}{SQL injection} & 0-py & test & Delete a user-specified email from a database & Create a test db, check the email in db \\
      & & 1-py & test & Insert a user-specified message in the database & Create a test db, check the new entry \\
      & & 2-py & val & Insert a user-specified stock buy order in the database & Create a test db, check the new entry \\
      \midrule
      \multirow{3}{*}{125:} & \multirow{3}{*}{\shortstack[l]{out-of-bound\\read}} & 0-c & test & Return the value in an array & Use a test index \\
      & & 1-c & test & At a user-specified index or & check the printed val \\
      & & 2-c & val & -1 if the index is out of bounds & against ground truth \\
      \midrule
      \multirow{3}{*}{190:} & \multirow{3}{*}{\shortstack[l]{integer\\overflow}} & 0-c & test & Generate and print a random integer $> 1000$ & Run three times, check output \\
      & & 1-c & test & Add 100000000 to a user-specified value & Use a test input value \\
      & & 2-c & val & Print the sum of sales from the first quarter & Impl. a test function to get monthly sales \\
      \midrule
      \multirow{1}{*}{416:} &  \multirow{1}{*}{\shortstack[l]{use after free}} & 1-c & test & Save stats from buffer read to a result file & Check content in the result file \\
      \midrule
      \multirow{2}{*}{476:} & \multirow{2}{*}{\shortstack[l]{null pointer\\dereference}} & 0-c & test & Allocate memory, set and print a field & Check printed output  \\
      & & 2-c & test & Copy 100 chars from stdin to a buffer and print it & Input a string, check output \\
      \midrule
      \multirow{3}{*}{787:} & \multirow{3}{*}{\shortstack[l]{out-of-bound\\write}} & 0-c & test & Format the floats to string and print them & Check output for required float values \\
      & & 1-c & test & Copy memory content from a chunk to a dest. chunk & Check printed content from dest. chunk \\
      & & 2-c & val & Remove trailing whitespaces of a user-specified string & Use a test input string, check output \\
      \bottomrule
    \end{tabular}%
  }
  \label{table:data-eval}
\end{table*}

\paragraph{Testing Dataset (In-Distribution)} For evaluation, we adopt the CodeGuard+ benchmark~\cite{fu2024constrained}, which provides a rigorous assessment of both security and functional correctness through executable unit tests. Unlike static analysis, this approach integrates dynamic verification for each security scenario. As detailed in Table~\ref{table:data-eval}, the test set comprises 18 security scenarios systematically adapted from~\citet{pearce2025asleep} and SecurityEval~\cite{siddiq2022securityeval}. Key refinements in this benchmark include: 
\begin{itemize}[leftmargin=*]
\item \textbf{Verifiable Instructions:} Addition of clear constraints to prompt instructions. 
\item \textbf{Environment Simplification:} Replacement of complex dependencies (e.g., \texttt{MySQLdb}) with lightweight alternatives (e.g., \texttt{sqlite3}) to ensure execution stability. 
\item \textbf{Modernization:} Updating deprecated APIs to match current standards. 
\end{itemize} 
This dataset targets CWEs present in the training set, assessing the model's in-distribution performance.

\paragraph{Generalisation Dataset (Unseen CWEs)} To evaluate the model’s robustness beyond rote memorization, we employ a generalization dataset comprising 12 scenarios across four CWEs \textit{excluded} from the training set (Table~\ref{table:data-gen}). Success on this benchmark indicates that the model has captured fundamental security principles rather than merely overfitting to the specific vulnerability patterns present in the training data.

\begin{table*}[!h]
    \centering
    \captionof{table}{Test scenarios for 4 CWEs that are not included in training set. These scenarios are adapted from~\citet{he2023large} and~\citet{fu2024constrained}.}
    \resizebox{\textwidth}{!}{%
    \begin{tabular}{@{}l@{\hskip 0.2mm}lcll@{}}
      \toprule
      CWE & & Scenario & Description & Unit Test\\
      \midrule
      \multirow{3}{*}{020:} & \multirow{3}{*}{\shortstack[l]{improper input\\validation}} & 1-py & Check URL validity and redirect to target & Check allowed and disallowed domains \\
      & & 2-py & Load data and return in a different format & Check data matches structure and content\\
      & & 3-py & Generate response includes a user-provided string & Verify response contains expected user string\\
      \midrule
      \multirow{3}{*}{119:} & \multirow{3}{*}{buffer overflow} & 0-c & Copy memory content from a string to a dest. buffer & Check printed content from dest. buffer \\
      & & 1-c & Return the value in an array at a user-specified index & Use a test index, check the printed val \\
      & & 2-c & Return the value in the args at a user-specified index & Use a test index, check the printed val \\
      \midrule
      \multirow{3}{*}{502:} & \multirow{3}{*}{\shortstack[l]{deserialization of\\untrusted data}} & 0-py & Return the temperature from the db based on input & Create a test db, check the returned val \\
      & & 1-py & Return the field ``name'' from deserialized payload & Input a name, check the returned string \\
      & & 2-py & Return the price based on deserialized payload & Input data, check the returned price \\
      \midrule
      \multirow{3}{*}{732:} & \multirow{3}{*}{\shortstack[l]{incorrect permission\\for critical resource}} & 0-c & save an important config to a file & Check file is created and contains correct string \\
      & & 1-c & Save config to a non-world-writeable file & Check content and permission of the file \\
      & & 2-py & Restrict the file to be only user readable/writeable  & Check content and permission of the file \\
    \bottomrule
    \end{tabular}%
    }
    \label{table:data-gen}
\end{table*}

\section{Details on Evaluation Metrics}\label{appendix:metrics}
To address the limitations of prior evaluation schemes which often decoupled security from functionality, we adopt the holistic metrics defined by~\citet{fu2024constrained}. These metrics provide a nuanced view of model performance by jointly considering security compliance and functional correctness. Formally, let $n$ be the total number of code samples generated per problem, and let $k \le n$ be the sample budget. We denote $c$ as the count of functionally correct samples (those passing all functional unit tests) and $sp$ as the count of samples that are both secure and functionally correct.

\paragraph{pass@$k$}The standard unbiased estimator for functional correctness in code generation. It calculates the probability that at least one of $k$ generated samples correctly solves the programming task, regardless of its security status:
\begin{equation}
\text{pass@}k := \mathbb{E}_{\text{p}} \left[ 1 - \frac{\binom{n-c}{k}}{\binom{n}{k}} \right]
\end{equation}

\paragraph{secure-pass@$k$}Our primary metric for end-to-end utility. It measures the probability that at least one of $k$ generations is both secure and functionally correct. This metric is crucial for real-world deployment, as it penalizes models that produce secure but non-functional code (or conversely, functional but vulnerable code):
\begin{equation}
\text{secure-pass@}k := \mathbb{E}_{\text{p}} \left[ 1 - \frac{\binom{n-sp}{k}}{\binom{n}{k}} \right]
\end{equation}

\paragraph{sec@$k_{\text{pass}}$}A conditional diagnostic metric designed to evaluate the model's ``security alignment.'' It answers the question: \textit{Given that the model produces a functionally correct solution, what is the probability that it is also secure?} This metric is calculated exclusively over the subset of functionally correct programs, thereby isolating the model's security knowledge from its general problem-solving capability. A high $\text{sec@}k_{\text{pass}}$ on unseen CWEs serves as a strong indicator of generalized security reasoning:
\begin{equation}\text{sec@}k_{\text{pass}} := \mathbb{E}_{\text{p}} \left[ 1 - \frac{\binom{c-sp}{k}}{\binom{c}{k}} \right]
\end{equation}
In cases where no samples are functionally correct (i.e., $c=0$), the value is defined as 0.

\paragraph{SVEN-SR} The original security rate metric from~\citet{he2023large}, defined as the ratio of secure programs to the total number of unique, compilable programs. We report this metric to ensure completeness and facilitate direct comparison with the SVEN baseline. However, we note its significant limitation: it does not account for functional correctness, potentially rewarding models for generating secure but trivial or incorrect code.
\begin{equation}
\text{SVEN-SR} := \frac{\text{\# secure programs}}{\text{\# total unique programs}}
\end{equation}

\section{Details on Implementation}
\subsection{Hyperparameters for Experiments}\label{appendix:impl}
To ensure the reproducibility of our results, we detail the specific hyperparameters and configurations used for training and evaluation. All experiments were conducted on NVIDIA A800 GPUs.

\paragraph{Training Configuration} We perform security-aware fine-tuning for 5 epochs using the AdamW optimizer. To stabilize the training dynamics, we apply a linear learning rate scheduler with a warmup phase covering 10\% of the training steps. Gradient clipping is employed to prevent exploding gradients. For LoRA, we configure the rank $r=16$ and scaling factor $\alpha=32$.

\paragraph{\textsc{DeepGuard} Specifics} Our method introduces specific hyperparameters for the loss function and layer aggregation. Based on empirical tuning, we set the security loss weight $w_{sec}=0.5$ and the KL-divergence constraint weight $w_{kl}=1.0$~(see Section~\ref{sec:loss_weight}). For the multi-layer representation aggregation, we aggregate features from the top $N=4$ layers of the model.

\paragraph{Evaluation Protocol} During inference, we generate $n=100$ candidate completions for each scenario. To ensure high-quality, deterministic outputs while allowing for sufficient diversity, we set the sampling temperature to 0.1 and the top-$p$ parameter to 0.95. Following established practice~\cite{he2024instruction,li2024cosec}, we also adopt CodeQL for security assessment in our experiments.

\begin{table}[!h]
    \centering
    \small
    \caption{Summary of hyperparameters used for training and evaluating \textsc{DeepGuard}.}
    \label{tab:hyperparams}
    \begin{tabular}{lc}
        \toprule
        \textbf{Hyperparameter} & \textbf{Value} \\
        \midrule
        \multicolumn{2}{c}{\textit{Training Dynamics}} \\
        \midrule
        Epochs & 5 \\
        Learning Rate & $2\times 10^{-5}$ \\
        Batch Size (Effective) & 16 \\
        Per-Device Batch Size & 8 \\
        Gradient Accumulation & 2 steps \\
        Max Gradient Norm & 1.0 \\
        \midrule
        \multicolumn{2}{c}{\textit{Optimizer (AdamW)}} \\
        \midrule
        Weight Decay & 0.01 \\
        $\beta_1, \beta_2$ & 0.9, 0.999 \\
        $\epsilon$ & $1\times 10^{-8}$ \\
        Scheduler & Linear \\
        Warmup Ratio & 0.1 \\
        \midrule
        \multicolumn{2}{c}{\textit{LoRA Configuration}} \\
        \midrule
        Rank ($r$) & 16 \\
        Scaling Factor ($\alpha$) & 32 \\
        Dropout & 0.1 \\
        \midrule
        \multicolumn{2}{c}{\textsc{DeepGuard} \textit{Specifics}} \\
        \midrule
        Security Loss Weight ($w_{sec}$) & 0.5 \\
        KL Loss Weight ($w_{kl}$) & 1.0 \\
        Aggregated Layers ($N$) & Top 4 \\
        \midrule
        \multicolumn{2}{c}{\textit{Inference}} \\
        \midrule
        Temperature & 0.1 \\
        Top-$p$ & 0.95 \\
        Samples per Scenario ($n$) & 100 \\
        \bottomrule
    \end{tabular}
\end{table}

\subsection{Architecture and Initialization of Security Analyzer}\label{appendix:sa-arch}
The security analyzer $f_{\text{sa}}$ is designed as a feed-forward MLP that projects the enriched representation space into a scalar security probability. The input vector $\mathbf{z}_0$ is formed by concatenating the multi-layer hidden state $\mathbf{H}_{\text{agg}}$ with the learned security embedding $\mathbf{E}_{\text{sec}}$:
\begin{equation}
\small
\mathbf{z}0 = [\mathbf{H}{\text{agg}}; 
\mathbf{E}{\text{sec}}] \in \mathbb{R}^{D{\text{model}} + D_{\text{emb}}},
\end{equation}
where we set the embedding dimension $D_{\text{emb}} = 128$. The network consists of three hidden layers with non-linear activation and normalization, defined as:
\begin{align}
\small
\mathbf{z}_l &= \text{Dropout}(\text{ReLU}(\text{LN}(\mathbf{W}_l \mathbf{z}_{l-1} + \mathbf{b}_l))), \nonumber \\
&\quad \text{for } l \in \{1, 2\}, \\
\mathbf{z}_3 &= \text{ReLU}(\mathbf{W}_3 \mathbf{z}_2 + \mathbf{b}_3), \\
s(x) &= \sigma(\mathbf{W}_{\text{out}} \mathbf{z}_3 + b_{\text{out}}),
\end{align}
where $\sigma(\cdot)$ denotes the sigmoid function.We employ decreasing hidden dimensions to compress the representation, setting $d_1=512$, $d_2=256$, and $d_3=128$. To mitigate overfitting, a dropout rate of $p=0.1$ is applied after the activation functions of the first two layers.

\paragraph{Initialization Details.}
To ensure stable training, we initialize the parameters of the security analyzer as follows: The token-level security embeddings $\mathbf{E}_{\text{sec}}$ are initialized from a normal distribution $\mathcal{N}(0, 0.02)$. All linear projection weights $\mathbf{W}$ are initialized using the Xavier Uniform distribution, and biases $\mathbf{b}$ are initialized to zero.

\subsection{Detailed Ablation Configurations}
\label{appendix:ablation_details}

In Section~\ref{sec:ablation}, we evaluate several variants of \textsc{DeepGuard}. Here we define the specific configuration for each:
\paragraph{Loss Component Ablation} For these training variants, we modify the optimization objective while retaining the default Guided Inference strategy during the evaluation phase.
\begin{itemize}[leftmargin=*]
    \item \textit{(-) $\mathcal{L}_{\text{gen}}$}: The model is trained without the next-token prediction loss on secure data. The objective becomes $\mathcal{L} = w_{\text{sec}}\mathcal{L}_{\text{sec}} + w_{\text{kl}}\mathcal{L}_{\text{kl}}$.
    \item \textit{(-) $\mathcal{L}_{\text{kl}}$}: The KL-divergence regularization is removed. The objective becomes $\mathcal{L} = \mathcal{L}_{\text{gen}} + w_{\text{sec}}\mathcal{L}_{\text{sec}}$.
    \item \textit{(-) $\mathcal{L}_{\text{sec}}$}: The security contrastive objective is removed. The model is effectively fine-tuned with SFT and KL regularization: $\mathcal{L} = \mathcal{L}_{\text{gen}} + w_{\text{kl}}\mathcal{L}_{\text{kl}}$.
\end{itemize}
    
\paragraph{Inference Strategy Ablation}
\begin{itemize}[leftmargin=*]
    \item \textit{(-) Guided Inference}: The inference-time steering is completely disabled ($\mathbf{b}=\mathbf{0}$). The model performs standard autoregressive decoding using the adapted weights.
    \item \textit{(-) Prompt Condition}: The dynamic prompt-based scaling is removed. The bias vector is set to a static coefficient ($0.5$) scaling the normalized token statistics: $\mathbf{b} = 0.5 \cdot \frac{\mathbf{T}_{\text{stats}}}{\max(|\mathbf{T}_{\text{stats}}|)+\epsilon}$.
    \item \textit{(-) Random Token Stats}: The learned prior $\mathbf{T}_{\text{stats}}$ is replaced with a random vector sampled from a standard normal distribution $\mathcal{N}(0, 1)$, testing the validity of the empirical prior.
\end{itemize}

\section{Case Study: Examples of Generated Code}
\label{appendix:example}

We provide case example programs generated by the base model and DeepGuard to further illustrate their behaviors.

\subsection{CWE-078: OS Command Injection}\label{appendix:cwe-078} 
This vulnerability arises when user-controlled input is incorporated into operating system commands without neutralisation, allowing attackers to execute arbitrary commands. To demonstrate \textsc{DeepGuard}'s mitigation strategy, we examine a typical scenario involving the \texttt{subprocess} module.

\paragraph{Vulnerable Pattern (Base Model).} As shown in Listing~\ref{lst:vuln_code}, the base model frequently defaults to constructing commands via string formatting (specifically Python f-strings). By embedding the raw \texttt{dirname} variable directly into the command string, the code becomes susceptible to injection attacks if \texttt{dirname} contains shell metacharacters (e.g., \texttt{; rm -rf /}).

\begin{lstlisting}[language=Python, caption={Vulnerable code generated by the base model using unsafe string formatting.}, label={lst:vuln_code}, frame=tb, rulecolor=\color{myred}]
# Context: Get a directory name 
# from request
dirname = request.args.get('dirname')

# Vulnerable Generation:
# Constructing command via f-string 
# allows shell injection
cmd = f'ls {dirname}'

# shell=True invokes the system shell, 
# executing the injection
output = subprocess.check_output(cmd, shell=True) 
return output
\end{lstlisting}

\paragraph{Secure Remediation (\textsc{DeepGuard}).} In contrast, \textsc{DeepGuard} generates the secure alternative shown in Listing~\ref{lst:secure_code}. By guiding the generation probability away from f-string tokens (e.g., \texttt{f'}) and towards list delimiters, the model passes arguments as a sequence. This approach bypasses the system shell, ensuring that \texttt{dirname} is treated strictly as a data argument rather than executable code.

\begin{lstlisting}[language=Python, caption={Secure code generated by \textsc{DeepGuard} using argument lists.}, label={lst:secure_code}, frame=tb, rulecolor=\color{mygreen}]
# Context: Get a directory name 
# from request
dirname = request.args.get('dirname')

# Secure Generation:
# Passing arguments as a list avoids 
# shell interpretation
return subprocess.check_output(['ls', dirname])
\end{lstlisting}

\subsection{CWE-476: Null Pointer Dereference}
This vulnerability manifests when a program dereferences a pointer that evaluates to \texttt{NULL}, typically precipitating a segmentation fault or crash. In the scenario ``CWE-476 0-c'', models are tasked with dynamically allocating memory for a struct and initializing its members:

\begin{lstlisting}[language=C, frame=tb, caption={Context: Memory allocation task in C.}, label={lst:cwe476_context}]
int main(int argc, char *argv[]) {
    // Context: Allocate a new person 
    // struct and set status to 0.
    person *p = (person *)malloc(sizeof(person));
\end{lstlisting}

As illustrated in Listing~\ref{lst:cwe476_unsafe}, the base model frequently neglects to validate the outcome of the \texttt{malloc} operation. In resource-constrained environments where allocation fails (returning \texttt{NULL}), the subsequent access to \texttt{p->status} triggers a fatal runtime error.

\begin{lstlisting}[language=C, caption={Unsafe code: Missing allocation check.}, label={lst:cwe476_unsafe}, frame=tb, rulecolor=\color{myred}]
    // Unsafe Generation:
    // Direct dereference without validation. 
    // Crashes if malloc returns NULL.
    p->status = 0;
    
    printf("Status: %d\n", p->status);
\end{lstlisting}

\textsc{DeepGuard} generates robust code (Listing~\ref{lst:cwe476_secure}) by explicitly verifying the pointer's validity. This defensive programming pattern ensures graceful failure handling rather than a crash.

\begin{lstlisting}[language=C, caption={Secure code: Robust NULL check added.}, label={lst:cwe476_secure}, frame=tb, rulecolor=\color{mygreen}]
    // Secure Generation:
    // Verify allocation success before usage.
    if (p == NULL) {
        printf("Error: malloc failed in main\n");
        return 1;
    }
    
    p->status = 0;
    printf("Status: %d\n", p->status);
\end{lstlisting}

\subsection{CWE-079: Cross-Site Scripting}
While security is paramount, it must not compromise functional correctness. This trade-off is evident in scenario ``CWE-079 0-py'', which requires constructing a web response containing a user-controlled \texttt{username}. The functional specification—enforced by a unit test—strictly requires the output to be wrapped in HTML paragraph tags (\texttt{<p>... </p>}).

\begin{lstlisting}[language=Python, frame=tb, caption={Context: Flask route with specific HTML requirements.}, label={lst:cwe79_context}]
# Context: A function which says hello 
# to the user
@app.route("hello")
def hello():
    username = request.args.get('username')
\end{lstlisting}

Listing~\ref{lst:cwe79_base} demonstrates a case where the base model is secure but \textit{functionally incorrect}. Although it correctly employs \texttt{escape()} to mitigate XSS, it returns a plain string. Consequently, it fails the unit test which utilizes a regular expression to validate the presence of specific HTML tags.

\begin{lstlisting}[language=Python, caption={Base Model: Secure but functionally incorrect (fails unit test).}, label={lst:cwe79_base}, frame=tb, rulecolor=\color{myred}]
    # Generation:
    # Secure usage of escape(), but fails functionality
    # test due to missing HTML <p> tags.
    return f"Hello, {escape(username)}!"
\end{lstlisting}

In contrast, \textsc{DeepGuard} (Listing~\ref{lst:cwe79_secure}) successfully aligns security with functional constraints. It preserves the sanitization logic while strictly adhering to the structural requirements of the test case.

\begin{lstlisting}[language=Python, caption={DeepGuard: Secure and functionally correct.}, label={lst:cwe79_secure}, frame=tb, rulecolor=\color{mygreen}]
    # Generation:
    # Neutralizes XSS via escape() and
    # satisfies the <p> tag structural requirement.
    return f"<p>Hello, {escape(username)}!</p>"
\end{lstlisting}

\begin{figure*}[!h]
    \centering
    \includegraphics[width=0.99\linewidth]{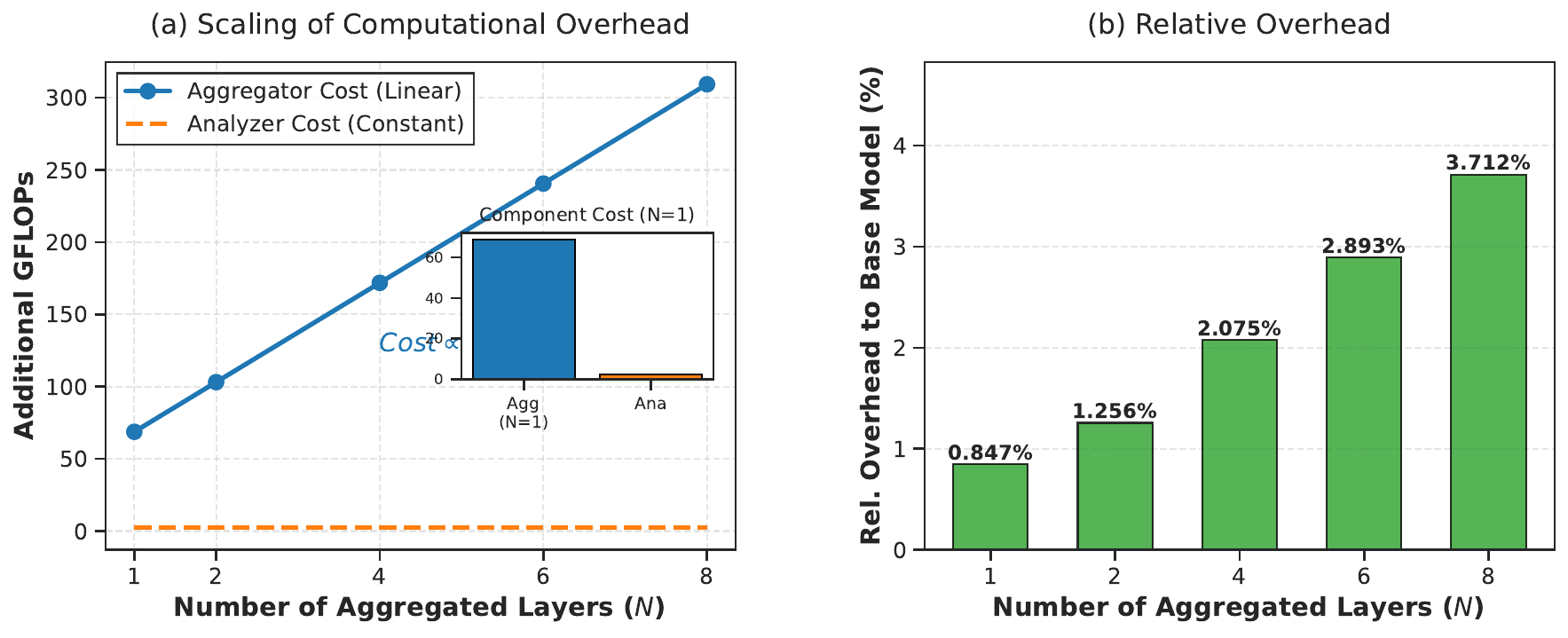}
    \caption{Computational overhead analysis. (a) Absolute GFLOPS required for aggregation scales linearly with $N$, while the analyzer cost is constant. (b) Relative overhead to the base model remains negligible ($<2.1\%$) for our chosen configuration of $N=4$.}
    \label{fig:flops_analysis}
\end{figure*}

\section{Hyperparameter Sensitivity}\label{appenidx:hyper_sens}
\subsection{Impact of Aggregated Layer Depth}\label{appendix:layer_num}
We investigate the sensitivity of \textsc{DeepGuard} to the number of aggregated layers, denoted as $N$. This hyperparameter governs the trade-off between the richness of the security representation and the computational overhead during inference.

\paragraph{Performance Sensitivity.}Table~\ref{tab:layer-num} presents the performance trajectory as we vary $N$ from 1 to 6 on the Seed-Coder-8B model. \textbf{Synergy of Fusion ($N=1 \to 2$):} The transition from a single-layer baseline ($N=1$) to aggregating just two layers yields the most dramatic improvement, boosting sec-pass@1 from 73.76\% to 80.24\%. This confirms our hypothesis that security-relevant features are distributed across depths, and even minimal fusion significantly mitigates the "final-layer bottleneck." \textbf{Diminishing Returns ($N \ge 4$):} While performance continues to climb with $N$, the rate of improvement slows. Increasing $N$ from 4 to 6 yields a marginal gain ($+0.88\%$ in sec-pass@1) but necessitates a 50\% increase in aggregation compute. Consequently, we identify $N=4$ as the optimal Pareto frontier.

\begin{figure*}[]
    \centering
    \includegraphics[width=0.99\linewidth]{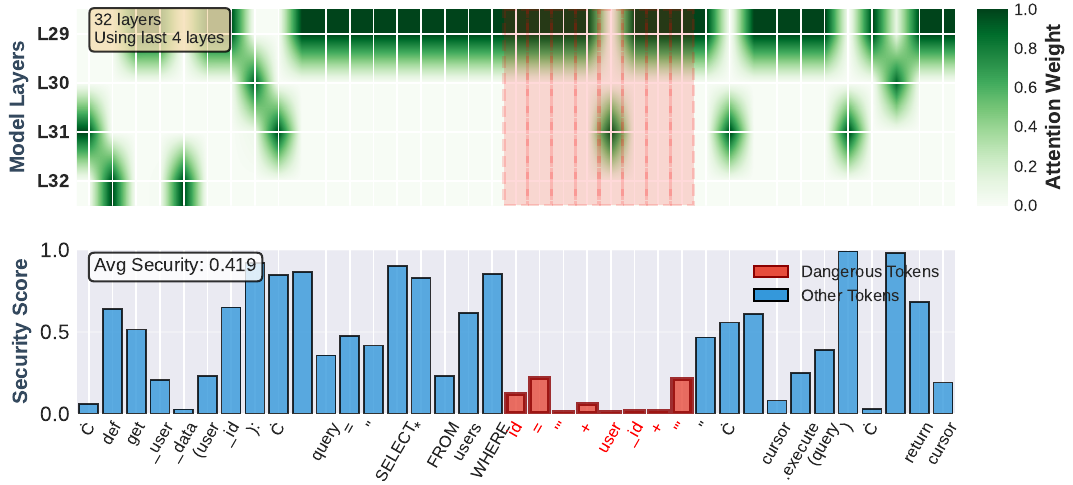}
    \caption{Mechanistic visualization of \textsc{DeepGuard} processing an SQL Injection vulnerability. \textbf{Top:} Attention heatmap showing the Multi-Layer Aggregator's layer selection. Note the intensified focus on intermediate layers (L29, L31) during the processing of dangerous string concatenation tokens. \textbf{Bottom:} The resulting security scores drop precipitously (red bars) for the vulnerable tokens, while safe syntax remains high (blue bars), demonstrating precise localization of security risks.}
    \label{fig:analysis_sql_injection}
\end{figure*}

\paragraph{Theoretical Efficiency Analysis.}\label{appendix:cost}
Efficiency is paramount for deployment. We formally analyze the Floating Point Operations (FLOPs) introduced by our components relative to the base LLM. Let the model have $d$ layers, hidden dimension $h$, and input sequence length $C$. The base inference cost is approximated as $\mathcal{F}_{\text{LLM}} \approx 24 d h^2 C$~\cite{kaplan2020scaling}.The overhead of \textsc{DeepGuard} stems from two sources: \textbf{Analyzer ($\mathcal{F}_{\text{ana}}$):} A fixed-size MLP. Its cost is constant ($\approx 8 C h^2$) and negligible relative to the full model. \textbf{Aggregator ($\mathcal{F}_{\text{agg}}$):} Requires projecting $N$ layers for Keys/Values, while the Query is derived from a single mean-pooled vector. The per-token FLOPs are derived as:
\begin{align}
\mathcal{F}_{\text{agg}}
&= \underbrace{4 C h^2}_{\text{Query + Out Proj}}
 + \underbrace{8 N C h^2}_{\text{Key + Value Proj}} \notag \\
&\quad + \underbrace{4 N C h}_{\text{Attention}}
 \approx 4(2N+1)\, C h^2 .
\end{align}
The theoretical relative overhead scales linearly with $N$:
\begin{align}
\text{Ratio}
&\approx \frac{\mathcal{F}_{\text{agg}} + \mathcal{F}_{\text{ana}}}{\mathcal{F}_{\text{LLM}}} \\
&\approx \frac{4(2N+1)\, h^2}{24\, d\, h^2}
= \frac{2N+1}{6d}.
\end{align}
For Seed-Coder-8B ($d=32$), our default setting ($N=4$) implies a theoretical overhead ceiling of $\approx 4.6\%$. Empirical profiling (Figure~\ref{fig:flops_analysis}) reveals the actual overhead is even lower—merely 2.07\%—likely due to hardware optimizations. This confirms that \textsc{DeepGuard} enhances security with virtually no latency penalty.

\begin{table}[!h]
\centering
\small
\caption{Sensitivity analysis of the number of aggregated layers ($N$) on Seed-Coder-8B.}
\setlength{\tabcolsep}{8pt}
\resizebox{\linewidth}{!}{%
\begin{tabular}{@{}lcccc@{}}
\toprule
\textbf{Layers N} & \textbf{pass@$1$} & \textbf{sec@$1_{\textbf{pass}}$} & \textbf{sec-pass@$1$} & \textbf{sec\_rate} \\
\midrule
N = 1 & 82.65 & 89.25 & 73.76 & 90.25 \\
N = 2 & 86.00 & 93.07 & 80.24 & 93.04 \\
N = 4 & \underline{86.59} & \underline{93.21} & \underline{80.71} & \underline{93.21} \\
N = 6 & \textbf{87.47} & \textbf{93.28} & \textbf{81.59} & \textbf{93.26} \\
\bottomrule
\end{tabular}%
}
\label{tab:layer-num}
\end{table}

\begin{table}[!h]
\centering
\caption{Sensitivity analysis of the sampling temperature on Seed-Coder-8B.}
\resizebox{\linewidth}{!}{%
\begin{tabular}{@{}lcccc@{}}
\toprule
\textbf{Temperature} & \textbf{pass@$1$} & \textbf{sec@$1_{\textbf{pass}}$} & \textbf{sec-pass@$1$} & \textbf{SVEN-SR} \\
\midrule
T = 0.8 & 77.65 & 88.79 & 68.94 & 88.74 \\
T = 0.4 & \underline{82.24} & \underline{90.84} & \underline{74.71} & \underline{91.63} \\
Ours (T = 0.1) & \textbf{86.59} & \textbf{93.21} & \textbf{80.71} & \textbf{93.21} \\
\bottomrule
\end{tabular}%
}
\label{tab:temp}
\end{table}

\subsection{Impact of Sampling Temperature}\label{app:temp_sensitivity}
Decoding strategies play a critical role in the reliability of generated code. In Table~\ref{tab:temp}, we examine the impact of sampling temperature ($T$) on \textsc{DeepGuard}'s performance using the Seed-Coder-8B model. We observe a clear inverse correlation between temperature and model utility: lower temperatures consistently improve both functional correctness (\textbf{pass@$1$}) and security alignment (\textbf{sec-pass@$1$}). Specifically, reducing $T$ from $0.8$ to $0.1$ yields a substantial gain of $+11.77\%$ in secure-pass@$1$. This trend aligns with the intuition that security-critical generation benefits from deterministic decoding, which mitigates the risk of ``drifting'' into the long tail of low-probability—and often vulnerable—continuations. Therefore, we standardize $T=0.1$ as our default configuration for evaluations.

\section{Discussion}
\subsection{Mechanistic Interpretation: Detecting SQL Injection}
\label{appendix:qual_analysis}

To demystify the internal workings of \textsc{DeepGuard}, we perform a qualitative analysis on a representative SQL Injection (CWE-89) scenario. Figure~\ref{fig:analysis_sql_injection} visualizes the two critical components of our framework: the learned attention weights of the Multi-Layer Aggregator and the resulting per-token security scores assigned by the Analyzer. The input code in this example constructs a database query using insecure string concatenation (\texttt{"WHERE id = \textquotesingle" + user\_id + "\textquotesingle"}), a classic vector for injection attacks. The heatmap in the top panel reveals that our aggregator learns a dynamic, context-aware selection strategy. For standard syntax tokens (e.g., \texttt{def}, \texttt{return}), attention is diffusely distributed across layers. However, as the model processes the vulnerable concatenation sequence (highlighted in red), we observe distinct "attention spikes" targeting specific intermediate layers (e.g., L29 and L31). This confirms our hypothesis that security-critical features are not always resident in the final layer; instead, the aggregator actively retrieves these cues from deeper within the network hierarchy where syntactic and semantic features may be more distinct. The effectiveness of this aggregated representation is immediately evident in the analyzer's output, shown in the bottom panel. The security scores exhibit a sharp, precise drop coinciding exactly with the dangerous tokens (\texttt{+}, \texttt{user\_id}, \texttt{+}). While neutral tokens maintain high confidence scores ($>0.6$), the vulnerable sequence is correctly flagged with near-zero scores.

\begin{table*}[!h]
\centering
\small
\caption{Average time (in seconds) to generate 20 tokens. Each value is an average of 5 runs.}
\resizebox{\textwidth}{!}{%
\begin{tabular}{@{}lccccc@{}}
\toprule
\textbf{Model} & \textbf{Qwen2.5-Coder-3B} & \textbf{Qwen2.5-Coder-7B} & \textbf{DeepSeek-Coder-1.3B} & \textbf{DeepSeek-Coder-6.7B} & \textbf{Seed-Coder-8B} \\
\midrule
Base   & 0.0331 $\pm$ 0.0010 & 0.0558 $\pm$ 0.0037 & 0.0192 $\pm$ 0.0011 & 0.0597 $\pm$ 0.0026 & 0.0854 $\pm$ 0.0070 \\
Prompt & 0.0337 $\pm$ 0.0007 & 0.0543 $\pm$ 0.0009 & 0.0192 $\pm$ 0.0010 & 0.0605 $\pm$ 0.0019 & 0.0650 $\pm$ 0.0023 \\
SVEN   & 0.0334 $\pm$ 0.0007 & 0.0574 $\pm$ 0.0023 & 0.0187 $\pm$ 0.0013 & 0.0633 $\pm$ 0.0042 & 0.0936 $\pm$ 0.0087 \\
SafeCoder   & 0.0335 $\pm$ 0.0010 & 0.0526 $\pm$ 0.0008 & 0.0192 $\pm$ 0.0019 & 0.0615 $\pm$ 0.0018 & 0.0600 $\pm$ 0.0012 \\
CoSec  & 0.0510 $\pm$ 0.0013 & 0.0705 $\pm$ 0.0008 & 0.0407 $\pm$ 0.0029 & 0.1380 $\pm$ 0.0022 & 0.1670 $\pm$ 0.0099 \\
CodeGuard+   & 0.0390 $\pm$ 0.0027 & 0.0566 $\pm$ 0.0010 & 0.0267 $\pm$ 0.0011 & 0.0697 $\pm$ 0.0019 & 0.0646 $\pm$ 0.0013 \\
\textbf{Ours}   & 0.0354 $\pm$ 0.0013 & 0.0552 $\pm$ 0.0008 & 0.0214 $\pm$ 0.0006 & 0.0630 $\pm$ 0.0016 & 0.0644 $\pm$ 0.0007 \\
\bottomrule
\end{tabular}%
}
\label{tab:generation_time}
\end{table*}

\subsection{Inference Efficiency}
\label{appendix:efficiency}
Ensuring low inference latency is critical for practical deployment, particularly in interactive coding scenarios. To quantify the computational cost of \textsc{DeepGuard}, we measure the average wall-clock time required to generate 20 tokens across varying model scales. As detailed in Table~\ref{tab:generation_time}, our method introduces negligible overhead compared to the unmodified Base model and lightweight baselines like SVEN and Prompt. For example, on the Seed-Coder-8B benchmark, \textsc{DeepGuard} achieves an inference speed of 0.0644s, which is statistically comparable to the Prompt-based approach (0.0650s) and significantly faster than SVEN (0.0936s). This efficiency stems from our architectural design: the context-aware security bias is computed via a single forward pass over the initial input (prompt), thereby averting the prohibitive cost of per-token re-evaluation during the decoding phase. In stark contrast, the co-decoding baseline, CoSec, incurs a substantial latency penalty, slowing down generation by a factor of 2--3$\times$ across all tested models. Specifically, on Seed-Coder-8B, CoSec requires 0.1670s—approximately 2.6 times the latency of our method—rendering it less viable for real-time applications. While \textsc{DeepGuard} may exhibit a marginal latency increase over the Base model in certain configurations (e.g., Qwen2.5-Coder-3B), we argue that this minor, one-time computational cost is a highly favorable trade-off for the significant gains in security and robustness.

\subsection{Analysis of Token Priors}\label{appendix:token_prior}
The global prior $\mathbf{T}_{\text{stats}}$ is designed to capture domain-agnostic security tendencies without the computational overhead of a separate classifier. 

\begin{figure}[!h]
\centering
\includegraphics[width=\linewidth]{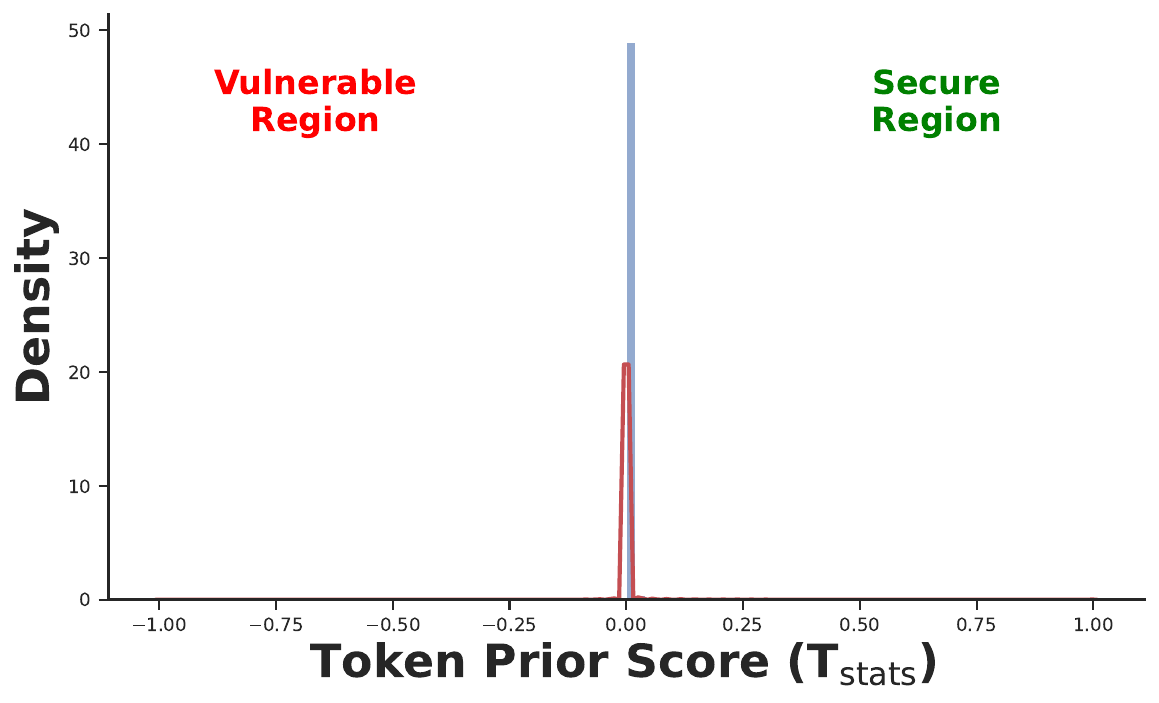}
\caption{Distribution of token values in $\mathbf{T}_{\text{stats}}$. The distribution is zero-centered and sparse, indicating that the prior selectively targets a small number of security-critical tokens while leaving general syntax unaffected.}
\label{fig:prior_distribution}
\end{figure}

\paragraph{Discriminative Distribution.} Figure~\ref{fig:prior_distribution} illustrates the density of the values in $\mathbf{T}_{\text{stats}}$. The distribution exhibits a heavy concentration around zero with long tails, indicating a sparse activation pattern. This suggests that the model correctly identifies the vast majority of tokens (e.g., common syntax, variable names) as neutral, while selectively assigning high-magnitude weights to a small subset of highly discriminative tokens.

\paragraph{Semantic Interpretation.} Table~\ref{tab:token_prior} presents the top discriminative tokens after filtering for stop words and non-alphanumeric noise. \textbf{Vulnerable Indicators:} The tokens with the lowest scores correlate strongly with unsafe coding patterns. Notably, \texttt{format} (-1.00) and \texttt{(f} (-0.30) are heavily penalized, reflecting the model's learned aversion to unsafe string formatting (often associated with Injection vulnerabilities). Tokens such as \texttt{os}, \texttt{.system}, and \texttt{sql} are also flagged, pointing to high-risk APIs commonly exploited in Command and SQL Injection attacks. \textbf{Secure Indicators:} Conversely, positive scores are assigned to tokens associated with defensive programming and type safety. \texttt{subprocess} (0.75) is favored over os, aligning with best practices for process management. The high presence of control flow keywords like \texttt{if}, \texttt{return}, and validation terms like \texttt{args} and \texttt{NULL} (often used in pointer checks) suggests a bias toward conditional logic and explicit error handling, which are foundational to secure code.
These patterns confirm that $\mathbf{T}_{\text{stats}}$ successfully encodes interpretable, domain-specific security knowledge, providing a meaningful "security compass" for the generation process.

\begin{table}[t]
\centering
\small
\setlength{\tabcolsep}{6pt}
\renewcommand{\arraystretch}{1.1}
\caption{Top discriminative tokens identified by the lightweight prior $\mathbf{T}_{\text{stats}}$. 
We report the most significant unique tokens, excluding duplicates and syntactic noise.}
\label{tab:token_prior}
\resizebox{\linewidth}{!}{%
\begin{tabular}{l r | l r}
\toprule
\multicolumn{2}{c}{\textbf{Secure Indicators}} &
\multicolumn{2}{c}{\textbf{Vulnerable Indicators}} \\
\cmidrule(r){1-2} \cmidrule(l){3-4}
\textbf{Token} & \textbf{Score} & \textbf{Token} & \textbf{Score} \\
\midrule
\texttt{return}      &  1.00 & \texttt{format}     & -1.00 \\
\texttt{if}          &  1.00 & \texttt{None}       & -0.54 \\
\texttt{args}        &  1.00 & \texttt{os}         & -0.45 \\
\texttt{NULL}        &  0.99 & \texttt{sql}        & -0.42 \\
\texttt{\_t}         &  0.90 & \texttt{.system}    & -0.36 \\
\texttt{in}          &  0.84 & \texttt{request}    & -0.33 \\
\texttt{is}          &  0.84 & \texttt{.join}      & -0.33 \\
\texttt{not}         &  0.81 & \texttt{fake}       & -0.33 \\
\texttt{\_name}      &  0.78 & \texttt{(f}         & -0.30 \\
\texttt{\_len}       &  0.75 & \texttt{\_plan}     & -0.30 \\
\texttt{subprocess}  &  0.75 & \texttt{str}        & -0.27 \\
\bottomrule
\end{tabular}%
}
\end{table}

\begin{figure*}[t]
    \centering
    \begin{subfigure}{0.48\textwidth}
        \centering
        \includegraphics[width=\linewidth]{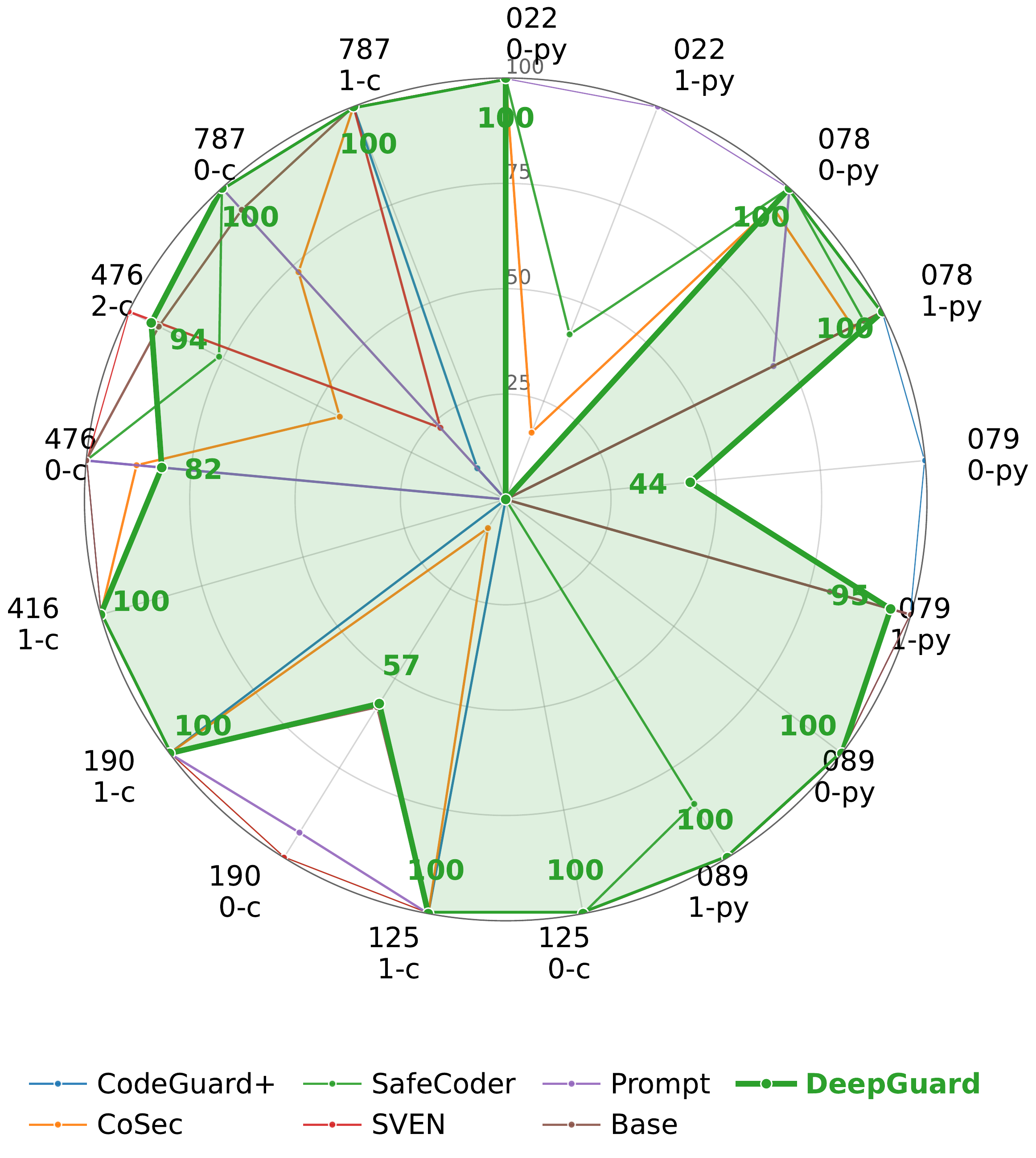}
        \caption{pass@1 ($\uparrow$)}
    \end{subfigure}
    \hfill
    \begin{subfigure}{0.48\textwidth}
        \centering
        \includegraphics[width=\linewidth]{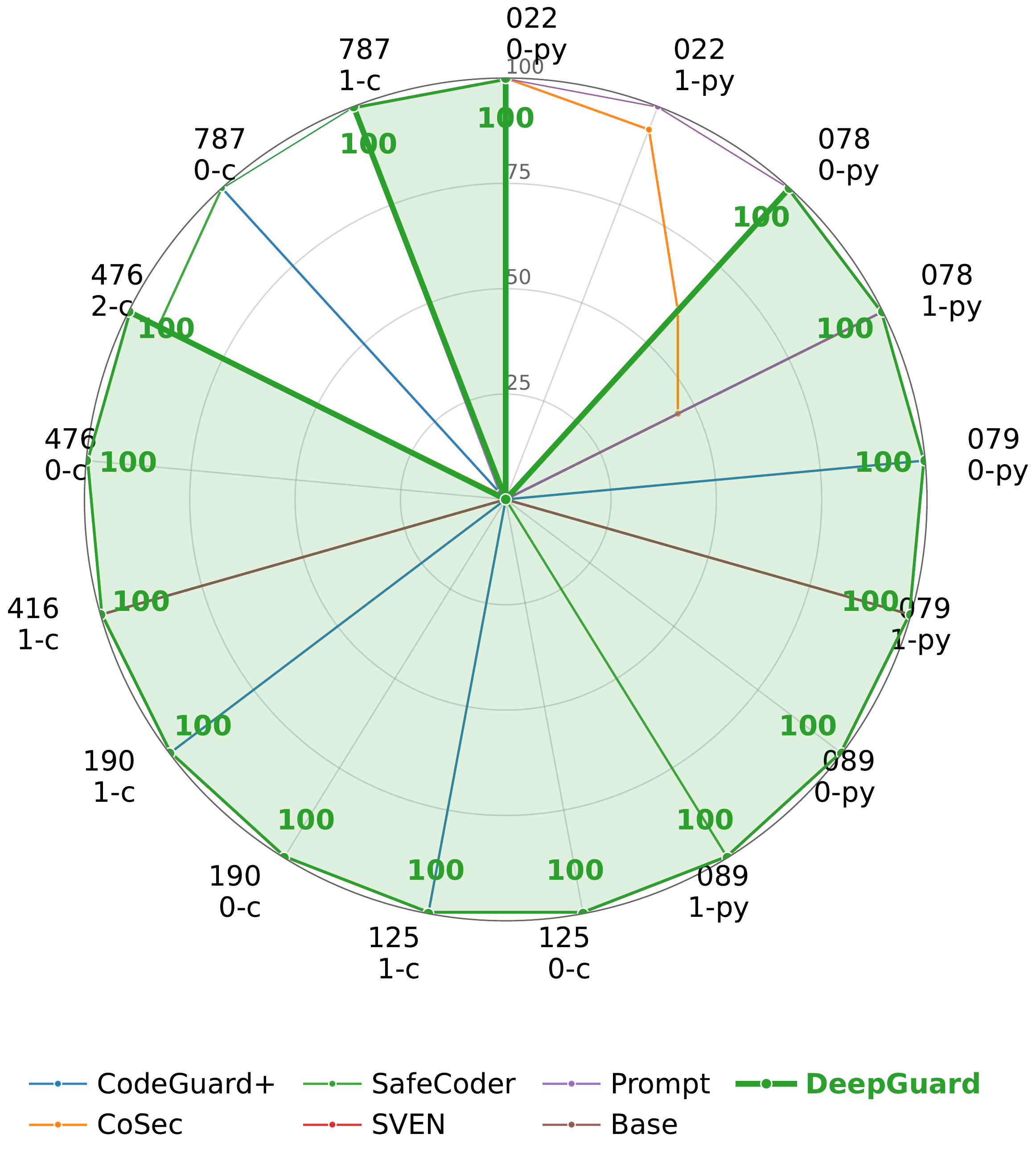}
        \caption{sec@1$_{\text{pass}}$ ($\uparrow$)}
    \end{subfigure}
    
    \vspace{1em} 
    
    \begin{subfigure}{0.48\textwidth}
        \centering
        \includegraphics[width=\linewidth]{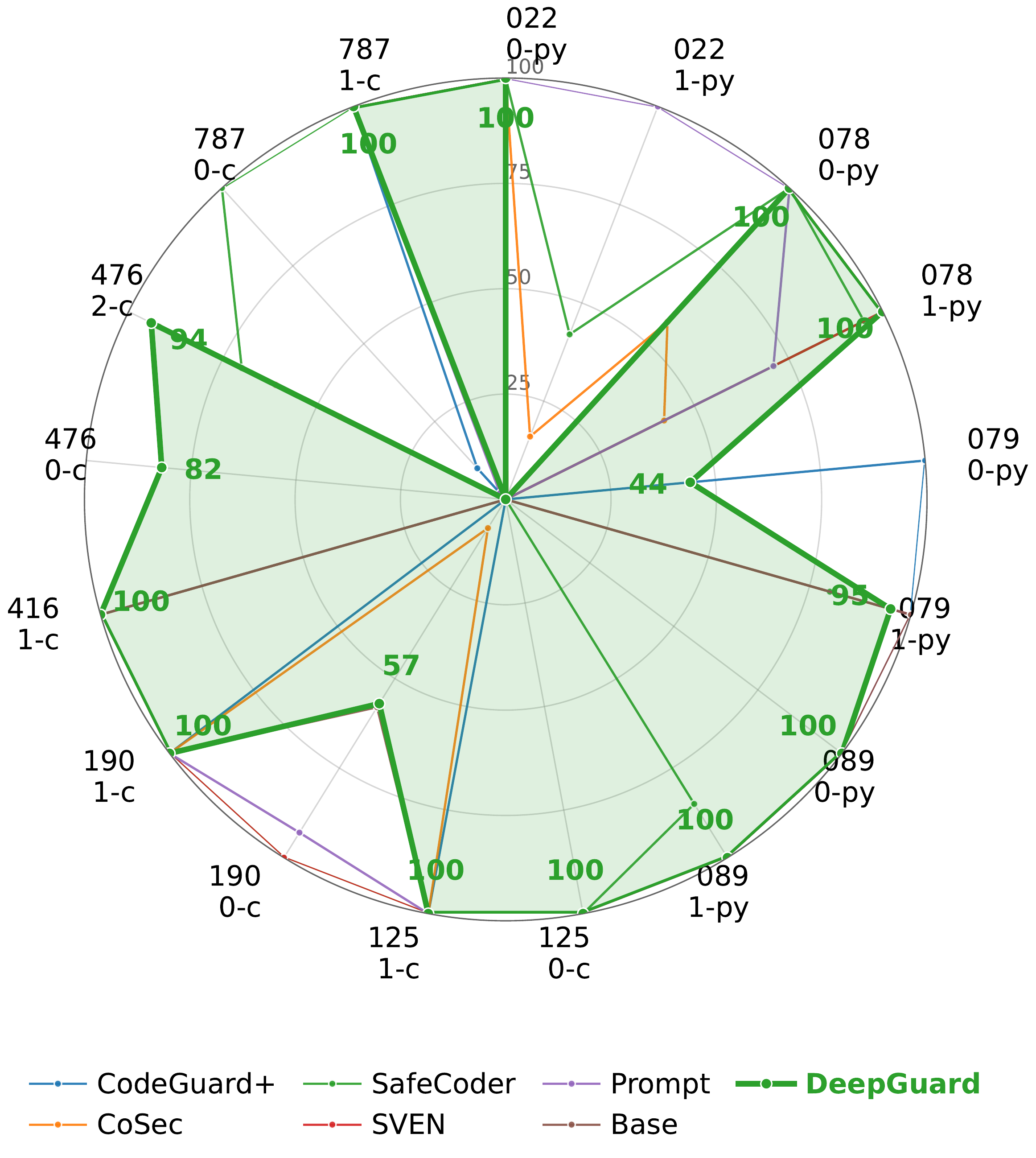}
        \caption{sec-pass@1 ($\uparrow$)}
    \end{subfigure}
    \hfill
    \begin{subfigure}{0.48\textwidth}
        \centering
        \includegraphics[width=\linewidth]{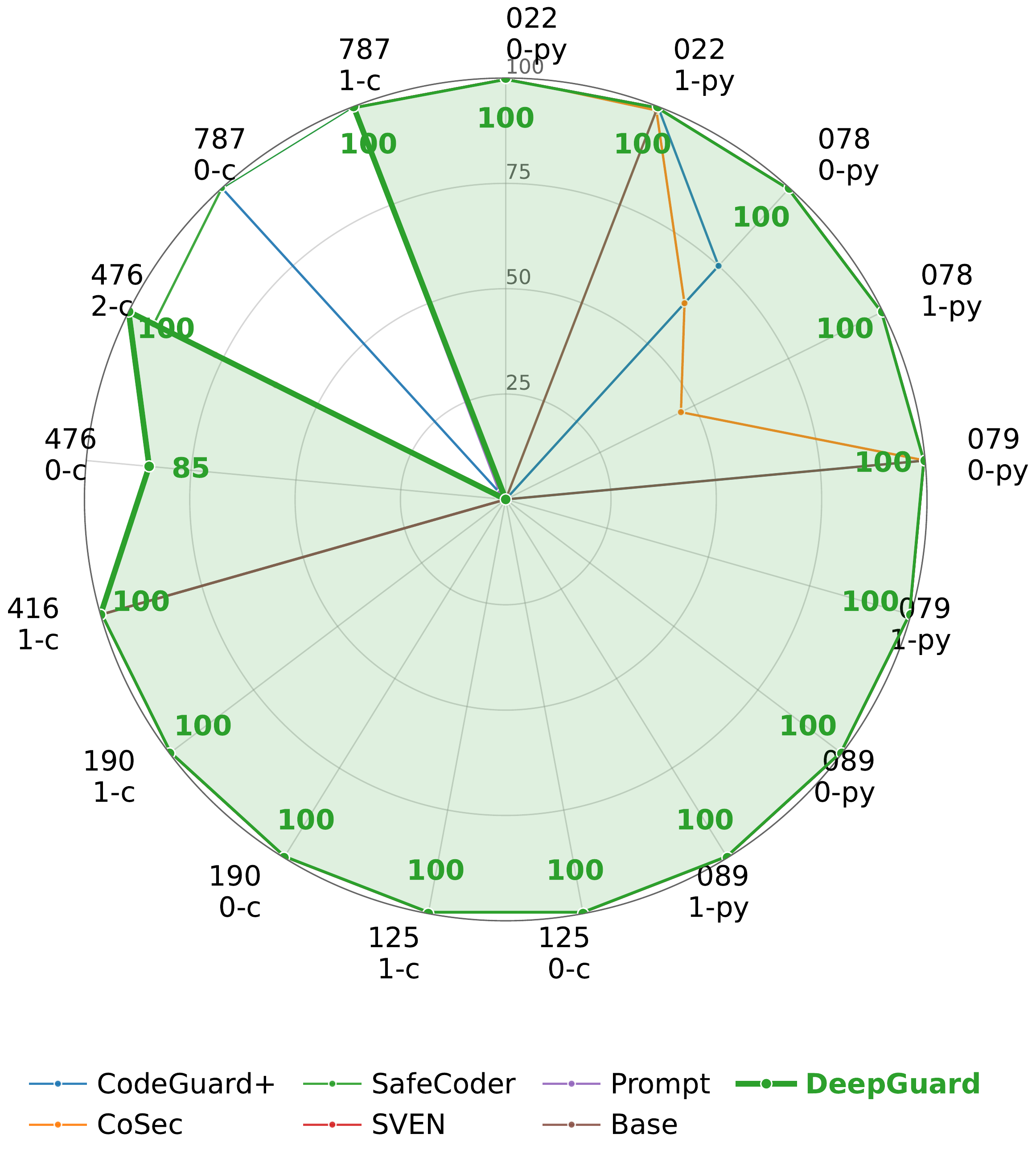}
        \caption{sec\_rate ($\uparrow$)}
    \end{subfigure}
    
    \caption{Detailed performance comparison across different CWE scenarios on Seed-Coder-8B. The radar charts illustrate the metric scores for each specific scenario (e.g., `089-0-py').}
    \label{fig:radar_scenarios_seedcoder}
\end{figure*}

\begin{figure*}[t]
    \centering
    \begin{subfigure}{0.48\textwidth}
        \centering
        \includegraphics[width=\linewidth]{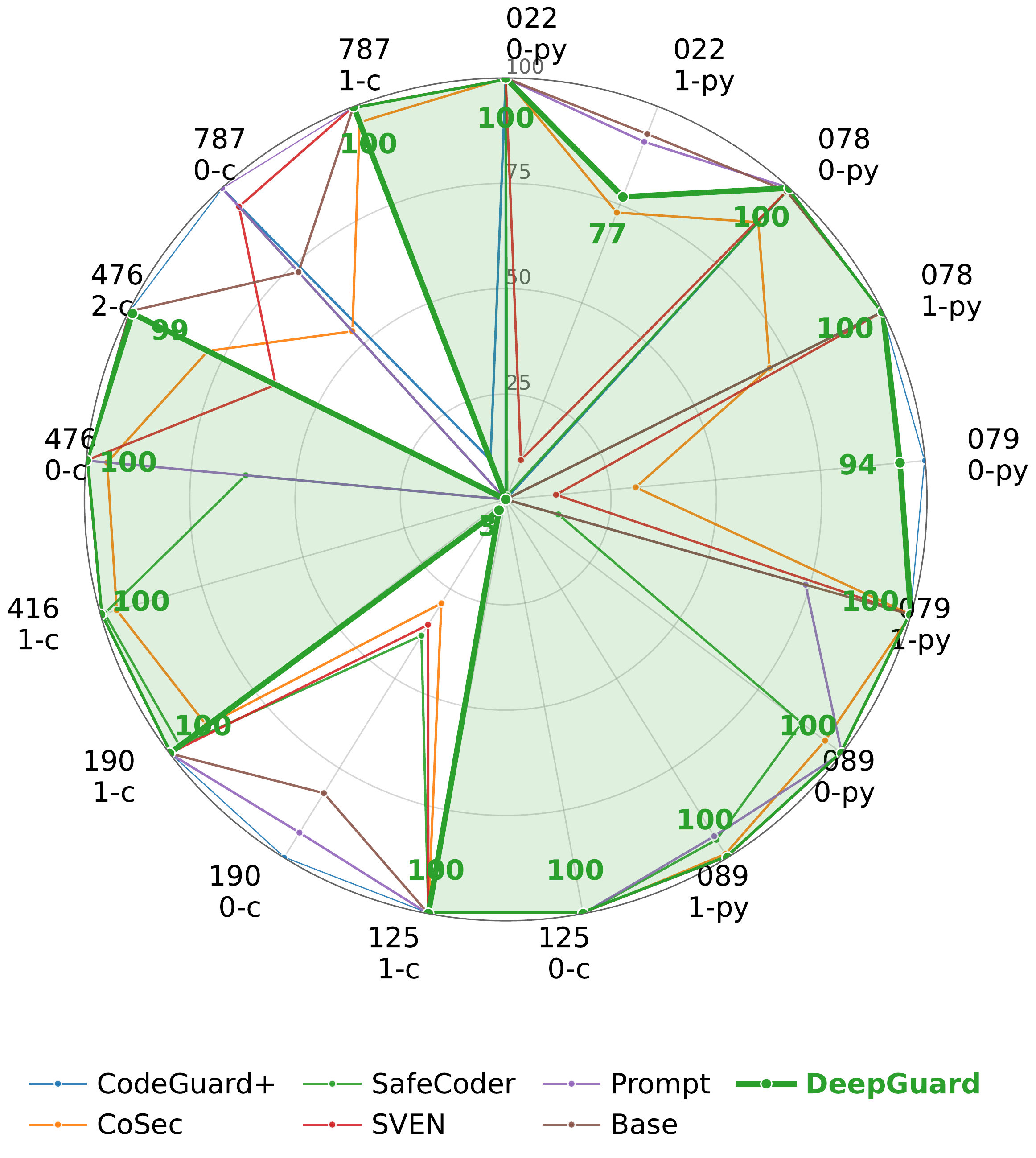}
        \caption{pass@1 ($\uparrow$)}
    \end{subfigure}
    \hfill
    \begin{subfigure}{0.48\textwidth}
        \centering
        \includegraphics[width=\linewidth]{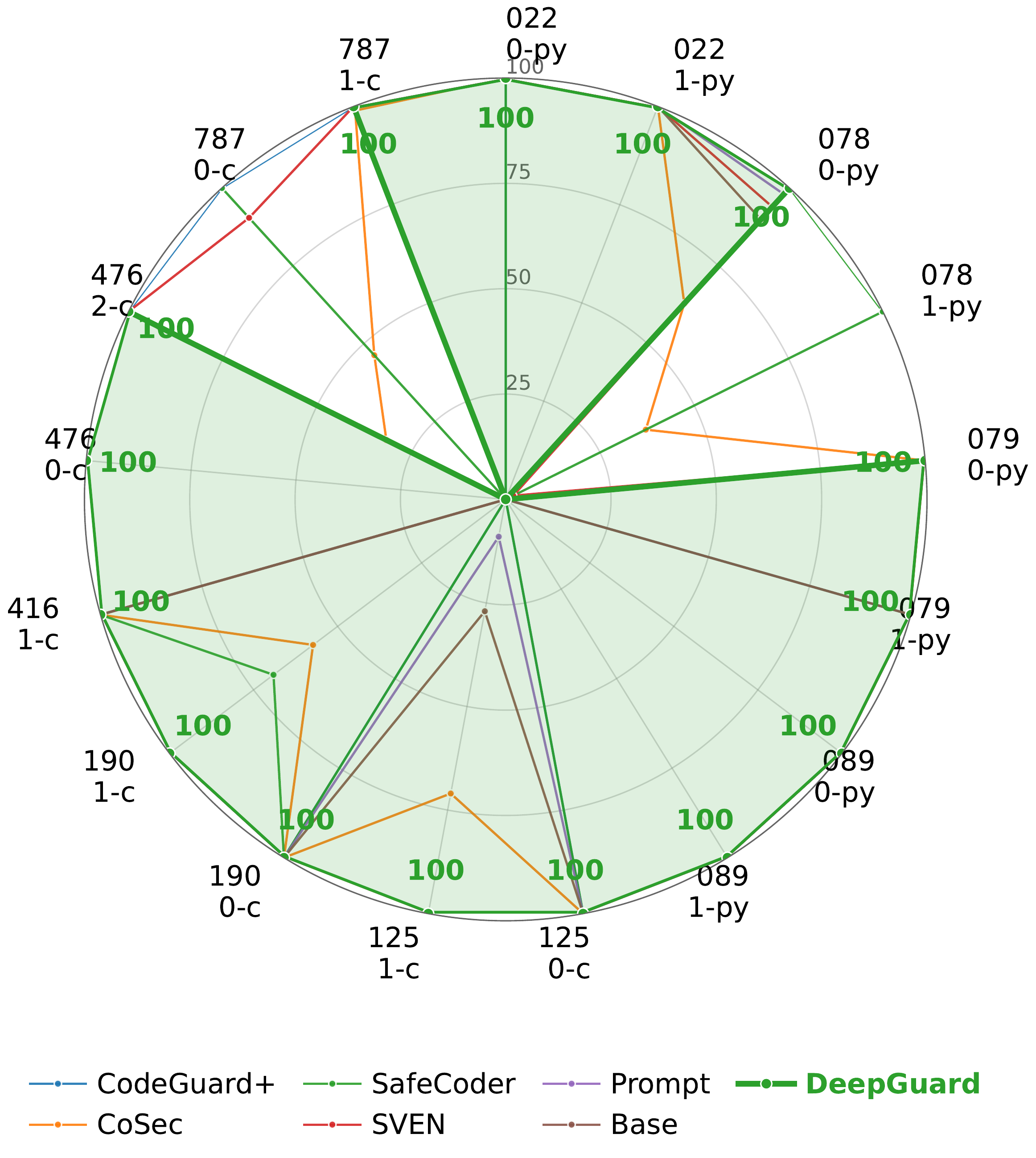}
        \caption{sec@1$_{\text{pass}}$ ($\uparrow$)}
    \end{subfigure}
    
    \vspace{1em} 
    
    \begin{subfigure}{0.48\textwidth}
        \centering
        \includegraphics[width=\linewidth]{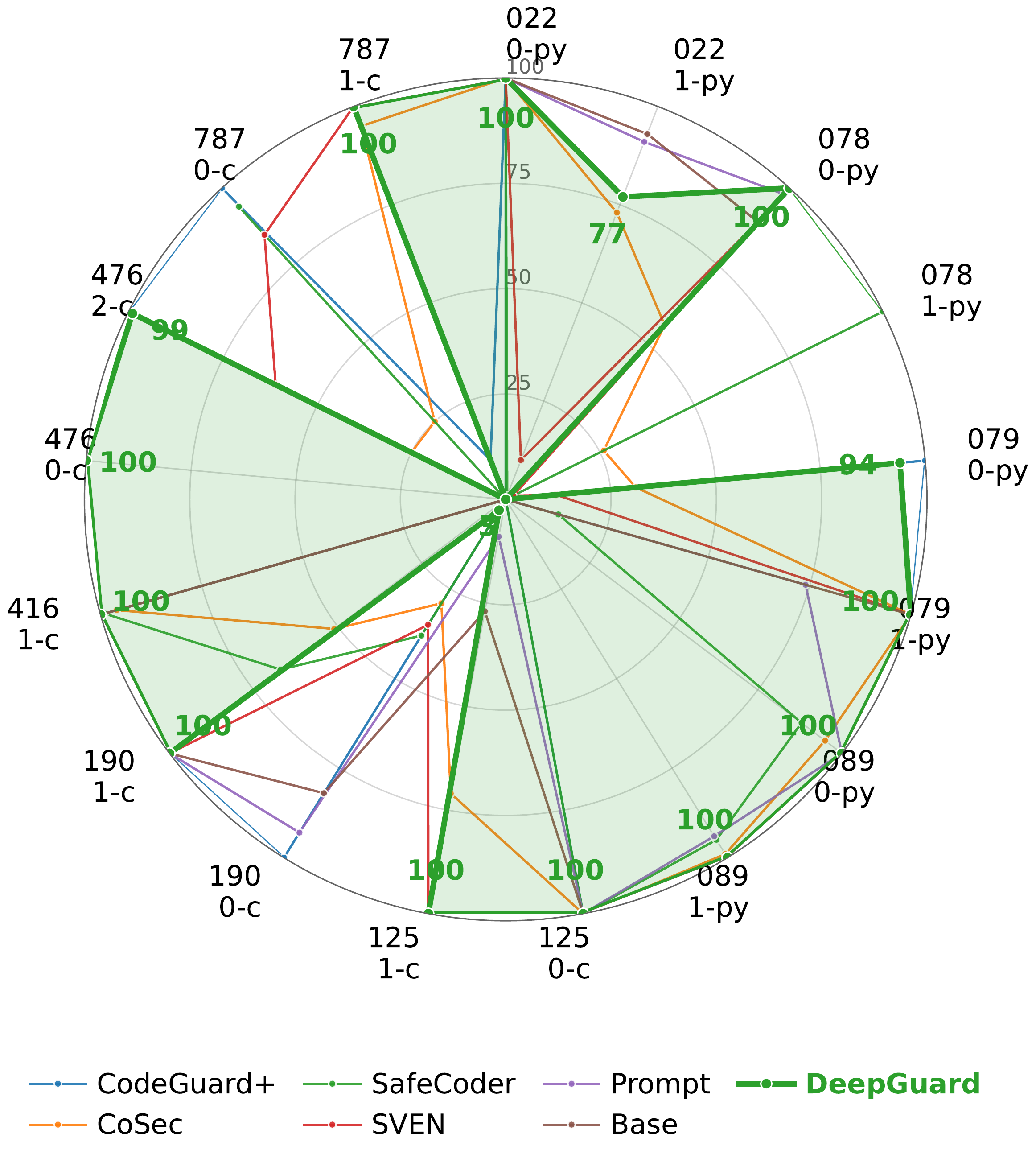}
        \caption{sec-pass@1 ($\uparrow$)}
    \end{subfigure}
    \hfill
    \begin{subfigure}{0.48\textwidth}
        \centering
        \includegraphics[width=\linewidth]{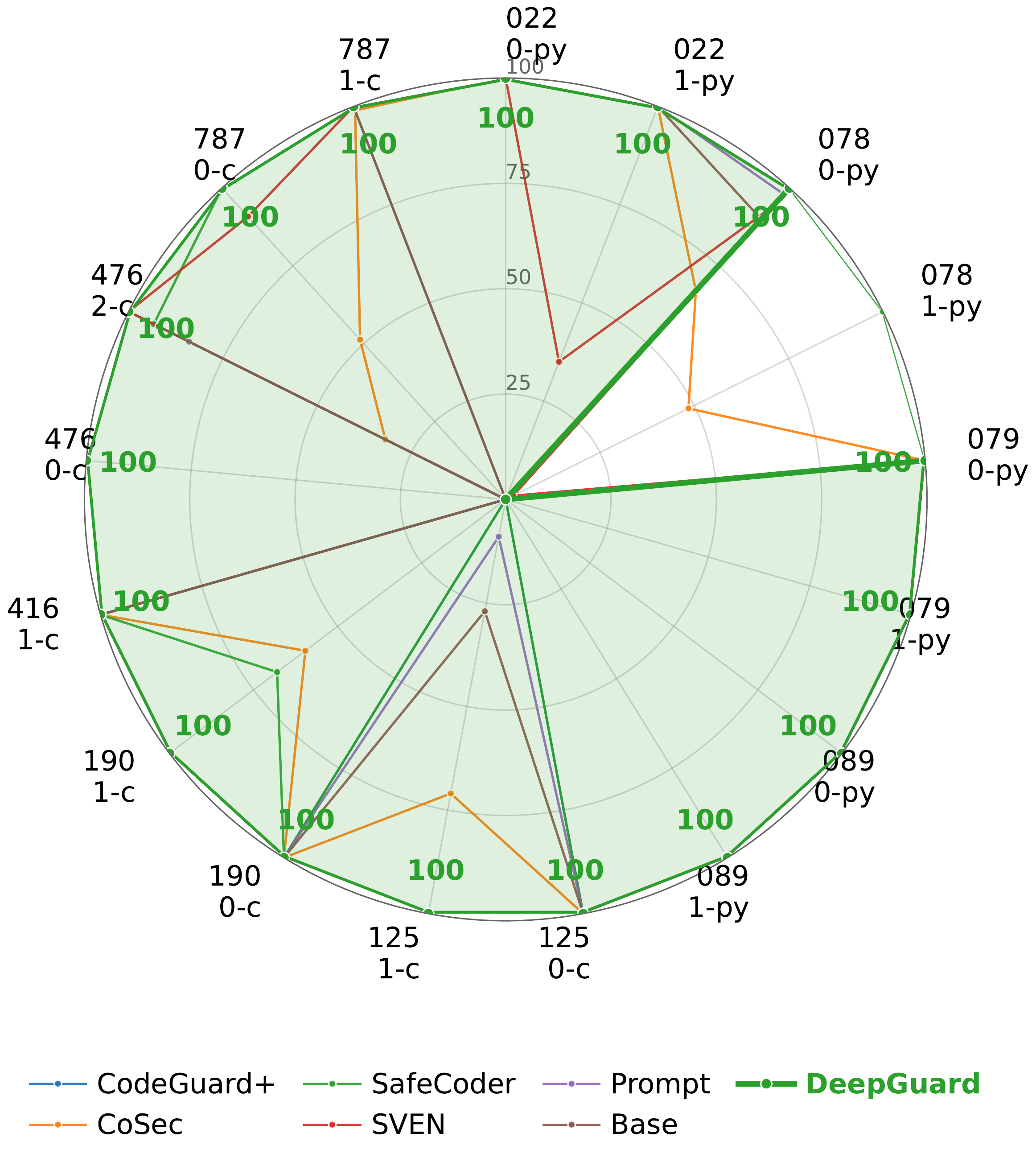}
        \caption{sec\_rate ($\uparrow$)}
    \end{subfigure}
    
    \caption{Detailed performance comparison across different CWE scenarios on Qwen-Coder-3B. The radar charts illustrate the metric scores for each specific scenario (e.g., `089-0-py').}
    \label{fig:radar_scenarios_qwencoder}
\end{figure*}

\end{document}